\documentclass[11pt,a4paper]{article}
\usepackage[left=2.35cm,top=3cm,bottom=3cm,right=2.3cm]{geometry}
\usepackage[english]{babel}
\usepackage[svgnames,x11names]{xcolor}
\usepackage{caption}
\usepackage[caption=false]{subfig}
\usepackage{graphicx}
\usepackage{tabularray}
\usepackage{makecell,multicol,floatrow,float}
\usepackage{amsmath}
\usepackage{amssymb}
\usepackage{slashed}
\usepackage{paralist}
\usepackage[T1]{fontenc}
\usepackage{cite}
\usepackage{authblk}

\usepackage[colorlinks=true
,urlcolor=blue
,anchorcolor=blue
,citecolor=blue
,filecolor=blue
,linkcolor=blue
,menucolor=blue
,linktocpage=true
,pdfproducer=medialab
]{hyperref}
\usepackage{orcidlink}
\definecolor{stuff}{rgb}{0.88,0.88,0.88}
\usepackage{tikz}
\usetikzlibrary{decorations.pathmorphing,decorations.markings,arrows.meta}
\tikzset{
    photon/.style={decorate, decoration={snake,amplitude=1.5pt,segment length=6pt}},
    tightphoton/.style={decorate, decoration={snake,amplitude=1.5pt,segment length=5pt}},
    fermion1/.style={draw=black, postaction={decorate},decoration={markings,mark=at position .54 with {\arrow{Latex[length=3pt]}}}},
	fermion2/.style={draw=black, postaction={decorate},decoration={markings,mark=at position .72 with {\arrow{Latex[length=3pt]}}}},
    photon1/.style={decorate, decoration={snake,amplitude=1.5pt,segment length=5.25pt}}
}
\def\centerarc[#1](#2,#3)(#4:#5:#6)% Syntax: [draw options] (center) (initial angle:final angle:radius)
    { \draw[#1] ({#2+#6*cos(#4)},{#3+#6*sin(#4)}) arc (#4:#5:#6); }

\numberwithin{equation}{section}

\renewcommand\Re{{\rm Re}}

\def\D{{\rm d}}
\def\cA{\mathcal{A}}
\def\cL{\mathcal{L}}
\def\cM{\mathcal{M}}
\def\cO{\mathcal{O}}
\def\cE{\mathcal{E}}

\newcommand\GeV{{\rm GeV}}
\newcommand\MeV{{\rm MeV}}
\newcommand{\Mew}{M_\text{EW}}
\newcommand\mcmule{{\sc McMule}}

\newcommand{\MS}{{\overline{\mathrm{MS}}}}

\newcommand{\mrad}{\mathrm{mrad}}

\DeclareMathOperator{\Real}{Re}

\makeatletter
\newcommand{\oset}[3][0ex]{%
	\mathrel{\mathop{#3}\limits^{
			\vbox to#1{\kern-2\ex@
				\hbox{$\scriptstyle#2$}\vss}}}}
\hypersetup{pdftitle={Parity violation in Moller scattering within low-energy effective field theory},pdfauthor={
Sophie Kollatzsch,
Daniel Moreno,
David Radic,
Adrian Signer}}
\makeatother

\author[1,2]{Sophie~Kollatzsch\,\orcidlink{0000-0002-8560-1619}\,}
\author[1]{Daniel~Moreno\,\orcidlink{0000-0001-8583-7632}\,}
\author[1,2]{David~Radic\,\orcidlink{0009-0001-2861-7032}\,}
\author[1,2]{Adrian~Signer\,\orcidlink{0000-0001-8488-7400}\,}
\affil[1]{PSI Center for Neutron and Muon Sciences, 5232 Villigen PSI, Switzerland}
\affil[2]{Physik-Institut, Universit\"at Z\"urich, 8057 Z\"urich, Switzerland}

\title{Parity violation in M\o{}ller scattering within \\[5pt]
low-energy effective field theory}
\date{}

\begin{document}

\begin{titlepage}
\clearpage\maketitle 
\thispagestyle{empty}

\begin{abstract}\noindent
We include electroweak effects in M\o{}ller scattering at low energies in an effective field theory approach and compute the left-right parity-violating asymmetry.
The calculation using low-energy effective field theory provides a solid framework to integrate out heavy particles with masses of the order of the electroweak scale, allowing the
resummation of all large logarithms between the electroweak scale and the scale, where QCD perturbation theory breaks down. The NLO electroweak corrections with leading logarithmic resummation, combined with QED corrections at NNLO and hadronic effects are implemented into the Monte Carlo framework \mcmule. Thus, we obtain a fully differential description and present results adapted to the MOLLER experiment. The potential impact of large logarithms at the next-to-leading logarithmic level is investigated. \\

\end{abstract}
\end{titlepage}

\section{Introduction} \label{sec:introduction}

The fundamentally different behaviour between the electromagnetic and the weak interactions concerning the conservation of parity
lead to the establishment of the electroweak (EW) sector of the Standard Model (SM) in 1978, after the
SLAC E122 experiment~\cite{Prescott:1978tm} observed parity violation due to weak neutral currents in the
deep inelastic scattering of polarised electrons from deuterium. Nowadays, the study of parity violation
remains an essential tool  for establishing the SM at the precision frontier and provides an
important testing ground for effects  beyond the SM (BSM).

In particular, the study of parity violation in electron scattering (PVES)~\cite{Souder:2015mlu}, i.e. of the asymmetry in the cross section of a left- or right-handed electron scattering off an unpolarised nuclear target,
provides an important tool for measuring the SM parameters, such as couplings to the $Z$ boson~\cite{Kumar:2013yoa},
to search for new neutral-current interactions~\cite{Ramsey-Musolf:1999qyv,Erler:2013xha,Davoudiasl:2023cnc}, and to measure properties of nuclei that are not accessible with other probes~\cite{Abrahamyan:2012gp,Erler:2014fqa,Cadeddu:2024baq}.

This has motivated the development of a large program of PVES experiments, especially at low energies, where an outstanding experimental accuracy can be achieved, allowing for precision tests of the SM and indirect searches of BSM physics.
Some of the most prominent PVES experiments are the SLAC E158~\cite{SLACE158:2005uay} and JLab MOLLER~\cite{MOLLER:2014iki} experiments, measuring the parity-violating asymmetry in M\o{}ller scattering, the JLab Qweak~\cite{Qweak:2014xey} and Mainz P2~\cite{Becker:2018ggl} experiments, measuring the asymmetry in electron-proton scattering, or the
SoLID experiment~\cite{Tian:2024xum}, planning to measure the asymmetry from electron scattering off a deuterium target
(see e.g.~\cite{Kumar:2013yoa} for a more detailed description of each experiment).
One of the main goals is to measure the weak mixing angle $s_{W}\equiv\sin\theta_W$ with outstanding precision.
%, as a tool to search for new physics effects.
So far, the extracted value of $s_{W}$ 
from low-energy experiments is not competitive with the measurement at the
$Z$ pole, where the two most precise measurements from SLAC~\cite{SLD:2000leq} and LEP1~\cite{ALEPH:2005ab} show some tension~\cite{Kumar:2013yoa}.
Nevertheless, the situation could change with the precision of the new-generation low-energy experiments, which could
help resolve this.

The PVES experiments are also complemented by other parity-violating experiments. The neutrino scattering experiments like
NuTeV provide an alternative method of measuring the interaction of the $Z$ boson with electrons and nuclei~\cite{Formaggio:2012cpf}.
Atomic parity violation experiments, like the measurement in Caesium~\cite{Wood:1997zq}, are based on the
observation of parity-violating interactions between the electrons and the nucleus in an atom, leading to transitions that are
otherwise strongly suppressed.

In this paper we will focus on the left-right asymmetry $A_{LR}$ in M\o{}ller scattering in the context of the upcoming MOLLER experiment~\cite{MOLLER:2014iki}, running at a centre-of-mass energy $\sqrt{s}\backsimeq0.1\,\GeV$. The experimental goal is to reach a precision of $2-3$\% in the parity-violating asymmetry. This would translate into a measurement of $s_{W}$ at the $0.1\%$ level.
Such an outstanding experimental precision must be accompanied
by equally precise theoretical predictions, whose improvement is the main goal of this paper.

The parity-violating asymmetry in M\o{}ller scattering at energies well below the EW
scale $\Mew$ is known up to one-loop order since 30 years ago~\cite{Derman:1979zc,Czarnecki:1995fw,Denner:1998um}. Also real radiation effects have been considered at next-to-leading order (NLO)~\cite{Petriello:2002wk,Zykunov:2004nk,Ilyichev:2005rx,Zykunov:2005md,Aleksejevs:2010ub,Aleksejevs:2010nf} and applied to the E158 and MOLLER experiments. 
Recently, two-loop corrections stemming from closed fermion loops, including the evaluation of nonperturbative hadronic
contributions, have also been obtained~\cite{Du:2019evk,Erler:2022ckm}. The evaluation of other 
subclasses of two-loop diagrams has been addressed in~\cite{Aleksejevs:2012xua,Aleksejevs:2013gxa,Aleksejevs:2015dba,Aleksejevs:2015zya,Aleksejevs:2018bzh}. However, the complete two-loop result is still missing. 
These results are based on SM calculations and often exploit the method of regions to obtain an expansion in the small scales, like the electron mass or centre-of-mass energy.

The one-loop corrections have been observed to reduce the tree-level asymmetry by $\sim 40\%$. The origin of the large corrections
is well understood. First, there is an accidental numerical suppression of the tree-level asymmetry which, due to the electron coupling to the $Z$ boson, is proportional to $1-4s_W^2$. At one loop, some contributions do not carry this tree-level suppression factor. In particular,  this is the case for quark loops, due to their different coupling to the $Z$ boson compared to the leptons. Furthermore, 
there are large logarithms induced by the hierarchy of the EW scale and the typical energy scale of
the process.
It was observed in~\cite{Czarnecki:1998xc,Czarnecki:2000ic} that the largest one-loop radiative corrections to
$A_{LR}$ can be captured through the definition of an effective parameter\footnote{Different definitions of $s_{W,\, \rm eff}^{2}$ exist in the
literature~\cite{Ferroglia:2003wa,Spiesberger:2024tfs}.} $s_{W,\, \rm eff}^{2}$, defined as
 \begin{eqnarray}
   s_{W,\, \rm eff}^{2}(Q^2) \equiv \kappa(Q^2) \hat{s}_{W}^{2}(M_Z)\,,
 \end{eqnarray}
where $Q^2$ is the momentum transfer, $\hat{s}_W$ is the weak mixing angle in the $\MS$ scheme,  and $\kappa(Q^2)$ includes in particular the $\gamma Z$ mixing effects.
In the literature,
this quantity is often referred to as the effective weak mixing angle, and its dependence on the momentum transfer as
running. This should not be confused with the actual renormalisation group running of the $\MS$ weak mixing angle $\hat{s}_{W}(\mu)$. In light of this observation, it has become standard to determine the
$Q^2$ dependence of the effective weak mixing angle, as a process-independent quantity,
useful to compare all low-energy parity-violating experiments, together with the extraction of $\hat{s}_{W}(M_Z)$~\cite{SLACE158:2005uay}. Therefore, the quantity $s_{W,\, \rm eff}$ has received a lot of attention and it has been improved beyond one-loop order by using the renormalisation group evolution (RGE) of $\hat{s}_{W}$~\cite{Erler:2004in}. The inclusion of nonperturbative effects in the low $Q^2$ range has also been refined by using lattice results and dispersion relations~\cite{Erler:2017knj,Erler:2024lds}. 

While $s_{W, \text{eff}}(Q^2)$ does capture some of the large logarithms, a full RGE-improved calculation is done most efficiently within an effective field theory (EFT) framework. The physics of low-energy M\o{}ller scattering is characterised by at least two
widely separated scales. The EW or hard scale $\mu_h$ and the kinematical or soft scale $\mu_s$, which is of the order of the typical centre-of-mass energy of the process. To fully exploit the hierarchy $\mu_s\ll\mu_h$ we will compute $A_{LR}$ in the low-energy effective field theory (LEFT)~\cite{Jenkins:2017jig}. The EW effects in LEFT are included through dimension~5 and dimension~6 operators. The Wilson coefficients of these operators are known through matching to the SM up to one loop~\cite{Dekens:2019ept} and the complete set of one-loop anomalous dimensions is also available~\cite{Jenkins:2017dyc}. There are tools~\cite{Aebischer:2018bkb,Fuentes-Martin:2020zaz} for the numerical evaluation of the RGE-improved Wilson coefficients at low scales. With this input, we can obtain a result that is accurate at NLO and resum all leading logarithms (LL) of the form $\ln(\mu_s/\mu_h)$. There has been substantial progress toward the computation of two-loop anomalous dimensions~\cite{Buras:2000if, Morell:2024aml, Naterop:2024cfx, Aebischer:2025hsx, Naterop:2025lzc, Naterop:2025cwg} which opens up the possibility to resum the logarithms at next-to-leading logarithmic (NLL) accuracy.

LEFT offers an efficient framework to combine EW corrections with QED effects. In our approach we compute the cross sections that enter $A_{LR}$ in a fully differential way. In particular, we also include real corrections. As we will argue, they play an important role and cannot be omitted for a direct comparison with experimental results. We base our calculations on the \mcmule{} framework~\cite{Banerjee:2020rww,McMule2022:manual} and deal with infrared (IR) divergences due to photon radiation with a well established subtraction method~\cite{Engel:2019nfw}.
Thus, in addition to NLO EW corrections through LL resummed Wilson coefficients, we take into account QED corrections up to next-to-next-to-leading order (NNLO). The latter are already available~\cite{Banerjee:2021qvi}.  Furthermore, the code is publicly available at
\begin{quote}
    \url{https://gitlab.com/mule-tools/mcmule}
\end{quote}
and can be used to produce results adapted to arbitrary experimental situations.

Our approach is sketched in Section~\ref{sec:framework} where we give more details on differences and similarities with earlier approaches in the literature. The details of the calculation are given in Section~\ref{sec:calculation}, where we first focus on the perturbative aspects within LEFT and then describe how we deal with the nonperturbative contributions for small centre-of-mass energies. In Section~\ref{sec:results} we present some results. We start in Section~\ref{sec:ResPerturbative} with the more conventional virtual results obtained in LEFT for energies that are sufficiently large to justify a purely perturbative treatment.
In this context, we also study the impact of NLL contributions. 
Results relevant for the MOLLER experiment are then shown in Section~\ref{sec:ResMoller}, where we also comment on the relevance of real corrections and NNLO QED effects. An outlook for future improvements in precision and extensions to other processes is given in Section~\ref{sec:conclusion}. Some more technical aspects are delegated to the Appendices~\ref{sec:appendixinput} and~\ref{sec:appendixexplicit}.

\section{Framework}\label{sec:framework}

Our aim is to provide a fully differential computation of M{\o}ller scattering at 
$\sqrt{s} \sim \mu_s \ll \Mew\sim \mu_h$. In accordance with the MOLLER experiment~\cite{MOLLER:2014iki} we assume longitudinally polarised incoming electrons scattering off unpolarised electrons at rest and take $\sqrt{s} = 106.03\,\MeV$ as our default value. This corresponds to a beam energy of $E_\text{beam}=11\,\GeV$.

In this section we describe our framework, sketch our approach for the computation of this process, and contrast it with previous approaches presented in the literature. The key difference is that we perform the computation strictly within LEFT, using the notation\footnote{As an exception to the rule, we denote the Wilson coefficients of LEFT by $C_i$ rather than $L_i$.} and conventions of~\cite{Jenkins:2017jig}. Our results are not formulated in terms of the weak mixing angle $s_W$, since such a quantity is not directly present in LEFT. Rather, it is expressed in terms of Wilson coefficients $C_i$.

The dominant effects in M{\o}ller scattering at low energies are due to QED. However, the parity-violating EW effects lead to a non-vanishing left-right asymmetry
\begin{align} \label{ALRdef}
  A_{LR} &\equiv \frac{\D\sigma_L-\D\sigma_R}{\D\sigma_L+\D\sigma_R}\, ,  
\end{align}
where $\D\sigma_{L/R}$ are the differential cross sections for left- or right-handed incoming electrons. Beyond leading order (LO), real and virtual corrections have to be taken into account.

Exploiting the hierarchy of scales $\mu_s\ll\Mew$, we use an EFT approach and work with the Lagrangian
\begin{align} \label{eq:Lagb}
  \cL_{\text{LEFT}}&= \cL_\text{QED} + \cL_\text{QCD} 
  + \sum_i C_i\, \mathcal{O}_i(\ell,q) \, ,
\end{align}
where $\mathcal{O}_i$ are dimension~5 and dimension~6 LEFT operators~\cite{Jenkins:2017jig}, written in terms of leptonic fields $\ell\in\{e,\mu,\tau\}$, light quark fields $q\in\{u,d,s,c,b\}$, and the photon and gluon field. For the evaluation of $\D\sigma_{L/R}$ at $\sqrt{s} \sim \mu_s$, the Wilson coefficients $C_i$ have to be evaluated at the (soft) scale of the process $\mu_s$. In the SM, the coefficients $C_i(\mu_s)$ are obtained by matching the SM to LEFT at the hard scale $\mu_h\sim\Mew$ and subsequent RGE to the soft scale. Through this procedure, all large logarithms of the form $\ln(\mu_s/\mu_h)$ are resummed. At higher orders in QED, there will still be logarithms of the form $\ln(m_e/\mu_s)$. This is a common feature of QED corrections to scattering cross sections and various strategies exist to resum them, see e.g.~\cite{WGRadCor:2010bjp,Aliberti:2024fpq}. In this paper, we do not perform such a resummation, but we do include fixed-order QED corrections up to NNLO.

In order to further illustrate our approach, we briefly discuss the calculation of the leading contribution to $A_{LR}$ and its $\mu$ dependence.  The leading result for $A_{LR}$ in LEFT reads
\begin{align}\label{ALRLO}
A_{LR} &= 
     -\frac{s\, y (1-y)}{(1-y+y^2)^2} 
    \frac{C_{ee,1111}^{V,LL}-C_{ee,1111}^{V,RR}}{2\pi \bar\alpha} + \ldots\, ,
\end{align}
where we have set the electron mass to zero, introduced the ratio of Mandelstam variables $y\equiv -t/s$, and denoted by $\bar\alpha$ the QED coupling in the $\MS$-scheme in LEFT. The use of this scheme for the QED coupling is tied to our aim of resumming all logarithms through the RGE, as discussed in more detail below. Using the Wilson coefficients obtained by matching the LEFT onto the SM at tree level and expressing them in terms of the Fermi constant $G_F$ and $\hat s_W$, we recover the well-known result
\begin{align}\label{ALRLOsw}
 A_{LR} &=  \frac{G_F}{\sqrt{2}}
    \frac{ s\, y (1-y)}{(1-y+y^2)^2}\,
    \frac{(1-4\,\hat s_W^2)}{2\pi \bar\alpha} + \ldots \, .
%     \textcolor{red}{\mbox{ {\bf DM}: check}}
\end{align}
Thus, in the EFT approach, the $\mu$ dependence of $A_{LR}$ is predominantly governed by the anomalous dimensions of $C_{ee,1111}^{V,LL}$ and $C_{ee,1111}^{V,RR}$ as well as the QED $\MS$ coupling. 
In the literature, the measurement of $A_{LR}$ is often phrased as a measurement of $\hat s_W$ and its running, in particular at low energies. Through the LO matching at $\mu_h=\Mew$ the two approaches are related as
\begin{align}\label{CLRmatching}
C_{ee,1111}^{V,LL}(\mu_h)-C_{ee,1111}^{V,RR}(\mu_h)& = 
%\frac{\pi \alpha}{\sqrt{2} M_W^2 s_W^2} \big(-1+4\,s_W^2\big) + \cO(\alpha) \\
-\frac{G_F}{\sqrt{2}} \big(1-4 \hat s_W^2(\mu_h)\big) + \cO(\alpha) \, ,
\end{align}
but the running of $C_{ee,1111}^{V,LL}-C_{ee,1111}^{V,RR}$ in LEFT differs from the running of $\hat s_W$. 
The precise evaluation of $\hat s_W$ and its RGE dependence has been studied in great detail in the literature~\cite{Erler:2014fqa, Erler:2022ckm,Erler:2004in}. The leading $\mu$ dependence is governed by
\begin{align}\label{RGEsw}
(4\pi)^2 \mu \frac{\D \hat s_W^2}{\D\mu} 
&= -\frac{2e^2}{3} \Big( n_e \big(1-4 \hat s_W^2\big) 
+ n_u \frac{2N_c}{9} \big(3-8 \hat s_W^2\big) 
+ n_d \frac{N_c}{9} \big(3 - 4 \hat s_W^2\big) \Big)\, ,
 \end{align}
where $n_e=3$, $n_u=2$ and $n_d=3$ are the number of active charged lepton and $u$- and $d$-type quarks, respectively, at $m_b<\mu < M_{\rm EW}$.
These terms also appear through the anomalous dimensions of $C_{ee,1111}^{V,LL}$ and $C_{ee,1111}^{V,RR}$. More precisely, it is the contribution of penguin diagrams with a closed fermion loop, shown as EW~3 in  Figure~\ref{fig:diag} that reproduces \eqref{RGEsw}. The corresponding logarithms can be absorbed in a process-independent manner, and their numerical impact has been shown to be very large, in particular for hadronic loops. However, there are additional leading logarithms that cannot be described through $\hat s_W(\mu_s)$. Indeed, the running of $C_{ee,1111}^{V,LL}$ and $C_{ee,1111}^{V,RR}$ also receives contributions from penguin diagrams with an open fermion line (also represented by EW~3), as well as diagrams with photonic corrections, represented by EW~2 of Figure~\ref{fig:diag}. Their equivalent in the full SM are vertex corrections and box diagrams involving heavy gauge bosons. 

Summarising, the leading $\mu$ dependence for $A_{LR}$ in LEFT  is governed by
\begin{align}
    (4\pi)^2 \mu \frac{\D}{\D\mu} & \Big( C_{ee,1111}^{V,LL}-C_{ee,1111}^{V,RR}\Big)
    =  
    12 e^2 Q_e^2 ( C_{ee,1111}^{V,LL} - C_{ee,1111}^{V,RR}) 
    \nonumber \\
    &
    + \frac{4}{3} e^2 Q_e \bigg( 
    Q_e \sum_{p=1}^3( 4 C_{ee,11pp}^{V,LL} + C_{ee,11pp}^{V,LR} - C_{ee,pp11}^{V,LR} - 4 C_{ee,11pp}^{V,RR}) 
    \nonumber
    \\
    &
    +
    N_c Q_d \sum_{p=1}^3( C_{ed,11pp}^{V,LL} - C_{de,pp11}^{V,LR} + C_{ed,11pp}^{V,LR} - C_{ed,11pp}^{V,RR}) 
    \nonumber
    \\
    &
    + N_c Q_u \sum_{p=1}^2( C_{eu,11pp}^{V,LL} - C_{ue,pp11}^{V,LR} + C_{eu,11pp}^{V,LR} - C_{eu,11pp}^{V,RR} ) \bigg)\,. \label{RGEleft}
\end{align}

Inserting the tree-level matching expressions for the Wilson coefficients on the r.h.s., 
the terms $Q_e Q_d$ and $Q_e Q_u$ agree with the $n_d$ and $n_u$ terms in~\eqref{RGEsw}. This also holds for the $Q_e^2$ terms due to the muon and tau, i.e. for $p\in\{2,3\}$, which agree with the corresponding $n_e$ terms associated to the muon and tau in~\eqref{RGEsw}. However, the $Q_e^2$ terms from electrons ($p=1$) differ. While these terms are proportional to the factor $1-4 s_W^2$ and, hence, are numerically suppressed, they contribute at leading logarithmic accuracy.  

Let us analyse how this difference in the RGE is translated into $A_{LR}$. 
The resummation of large logarithms from the RGE of $\hat s_W$ leads to the following result for $A_{LR}$ at LL accuracy
\begin{eqnarray}
 A_{LR}^{\text{run}\,\hat{s}_W} =
 \frac{G_F}{\sqrt{2}}  \frac{ s y (1-y) }{(1 - y + y^2) ^2 } \frac{1}{2\pi \bar\alpha}
  \bigg[ (1-4\hat s_W^2) + \frac{20N_c}{27}\frac{\bar\alpha}{\pi}L
  + \mathcal{O}(\alpha^2 L^2)\bigg]\,,
\end{eqnarray}
where $L\equiv \ln(\mu_s/\mu_h)$ and we only show explicitly the single logarithm term. 
We observe that the $n_e$ terms cancel out between the running of $s_W$ and the electromagnetic coupling, so terms coming from closed lepton loops do not contribute. In other words, the $\alpha L$ term is proportional to $N_c$ only. In the EFT we obtain
\begin{align}
 A_{LR} &=
 \frac{G_F}{\sqrt{2}}  \frac{ s y (1-y) }{(1 - y + y^2) ^2 } \frac{1}{2\pi \bar\alpha}
  \bigg[ (1-4\hat s_W^2) + \bigg(\frac{20N_c}{27} + \frac{11}{3}Q_e^2 (1 - 4\hat s_W^2)\bigg)\frac{\bar\alpha}{\pi}L + \mathcal{O}(\alpha^2 L^2)\bigg]
  \nonumber
  \\
  &=
  \frac{G_F}{\sqrt{2}} \frac{ s y (1-y) }{(1 - y + y^2) ^2 } \frac{1}{2\pi \bar\alpha}
  \bigg[ (1-4\hat s_W^2) + \bigg(\frac{53}{9} - \frac{44}{3}\hat s_W^2\bigg)\frac{\bar\alpha}{\pi}L + \mathcal{O}(\alpha^2 L^2)\bigg]\,,
\end{align}
where the difference compared to the previous result is proportional to 
the small factor $1-4\hat s_W^2$ and to $Q_e^2$, and it is coming again from penguin diagrams with an open fermion line EW~3 and diagrams with photonic corrections 
EW~2. This term is missing in the SM approach. 
The missing term makes the coefficient in front of the $\alpha L$ about $13\%$ larger in the EFT than in the SM calculation. Again, we observe that the terms coming from closed lepton loops do not enter, and the terms coming from closed quark loops coincide with the ones obtained from the running of $\hat s_W$.

It has long been noted that the use of a running $\hat s_W(\mu_s)$ does not reproduce all leading logarithms~\cite{Erler:2004in} of M{\o}ller scattering and that there are additional process-dependent logarithms. With LEFT we have a framework that allows for the resummation of all large logarithms of the form $\ln(\mu_s/\mu_h)$. The price we have to pay is the restricted range of applicability, since we need $\sqrt{s} \ll \Mew$. At tree level, a calculation reveals that the difference between the full SM result and the LEFT result for $A_{LR}$ decreases from approximately 5\% at $\sqrt{s}=30\,\GeV$ to 0.5\% at $\sqrt{s}=5\,\GeV$. Of course, it is trivial to add the tree-level result in the full SM. From an EFT point of view this would correspond to adding terms with dimension higher than 6. However, at higher order, the EFT restricted to dimension~6 offers an efficient and systematic way to include EW corrections. 

For the inclusion of real emission of photons and their combination with virtual corrections within the LEFT framework, it is possible to use standard techniques from higher-order QED calculations. 
We need to include contributions with one (at NLO) or two (at NNLO) additional photons in the final state. This allows to combine virtual and real corrections and cancel IR singularities in the usual way. As we will argue this is essential to obtain results directly applicable to a realistic experimental situation. 

In a proper EFT calculation, at $\mu\sim 5\,\GeV$ the $b$ quark is integrated out and LEFT$_{f=7}$ (without a $b$ quark) is matched to LEFT$_{f=8}$ (with a $b$ quark). In addition to threshold corrections, this also affects the anomalous dimensions. In principle, this step has to be repeated at each threshold. However, for scales approaching $\mu\sim 2\,\GeV$ we enter the nonperturbative domain of the strong interactions, where light quarks (and gluons) are no longer the appropriate degrees of freedom.

\section{Calculation}\label{sec:calculation}

In this section we will give some technical details of the computation. In Section~\ref{sec:perturbative} we start with the region $\sqrt{s}\gtrsim 5\,\GeV$, where we can rely on a perturbative approach. Section~\ref{sec:dirtyWorld} describes our treatment of the nonperturbative effects at smaller values of $\sqrt{s}$.

\subsection{Perturbative calculations in LEFT} \label{sec:perturbative}

In order to systematically combine the resummation of large logarithms $\ln(\mu_s/\mu_h)$ with fixed-order calculations, in what follows, we will discuss how to obtain for $A_{LR}$ an RGE-improved result at NLO in LEFT and NNLO in QED. This allows us to obtain an expression for $A_{LR}(s)$ for values of $\sqrt{s}$ that are small enough to justify the application of LEFT. In this section, we will also assume that $\sqrt{s}$ is sufficiently large to avoid complications from the nonperturbative regime of QCD.

The perturbative expansion of the cross sections that enter \eqref{ALRdef} is written as
\begin{align} \label{dsigmaPT}
    \D\sigma_{L/R}&=
     \sum_{i=0}^{\infty}\alpha^{i+2}\,\D\sigma^{(i)}
     + \sum_{j=0}^\infty \lambda\,\alpha^{j+1}\,\D\sigma_{L/R}^{(j,1)}
     + \cO(\lambda^2)\, ,
\end{align}
where we restrict ourselves to at most a single insertion of a higher-dimensional operator. These insertions are labelled with a generic bookkeeping factor $\lambda$. For the combination of QED and EW corrections we apply a naive counting scheme $\lambda \sim \alpha^2$. Thus, we include terms with $i\le 2$ and $j\le 1$ in \eqref{dsigmaPT}. Since QED is parity conserving, the numerator of \eqref{ALRdef} vanishes in pure QED. However, the denominator of \eqref{ALRdef} is affected by the pure QED corrections $\D\sigma^{(i)}$.

Once the cross sections \eqref{dsigmaPT} are known, they can be directly inserted into \eqref{ALRdef} to evaluate what we call the `unexpanded' or `full' result for $A_{LR}$. The corresponding LO and NLO results read
\begin{subequations}\label{ALRfull}
\begin{align} \label{ALRfullLO}
 A_{LR}^{\text{LO full}} &= \frac{\lambda\,\D\sigma_{L-R}^{(0,1)}}{2(\alpha\, \D\sigma^{(0)} + \alpha^2\, \D\sigma^{(1)}) + \lambda\, \D\sigma_{L+R}^{(0,1)}} \, ,
 \\
\label{ALRfullNLO}
 A_{LR}^{\text{NLO full}} &= 
 \frac{\lambda\,\D\sigma_{L-R}^{(0,1)} + \alpha \lambda\,\D\sigma_{L-R}^{(1,1)}}{2 (\alpha\,\D\sigma^{(0)} + \alpha^2 \D\sigma^{(1)}+  \alpha^3\,\D\sigma^{(2)})  + \lambda\,\D\sigma_{L+R}^{(0,1)} + \alpha\lambda\, \D\sigma_{L+R}^{(1,1)}} \, ,
\end{align}
\end{subequations}
where we have defined $\D\sigma_{L\pm R}^{(j,1)}\equiv \D\sigma_L^{(j,1)}\pm\D\sigma_R^{(j,1)}$. 
Alternatively, \eqref{ALRdef} can be expanded. The presence of two small scales, $\lambda$ and $\alpha$, however leads to ambiguities. A similar situation was encountered in top-quark asymmetries when including EW and QCD corrections~\cite{Czakon:2017lgo}. We define the `expanded' version of $A_{LR}$ as 
\begin{subequations}\label{ALRexp}
\begin{align} \label{ALRexpLO}
 A_{LR}^{\text{LO exp}} &= 
 \frac{\lambda}{\alpha} \frac{\D\sigma_{L-R}^{(0,1)}}{2\,\D\sigma^{(0)}}\,, 
   \\ \label{ALRexpNLO} 
A_{LR}^{\text{NLO exp}} &= 
 \frac{\lambda}{\alpha} \frac{\D\sigma_{L-R}^{(0,1)}}{2\,\D\sigma^{(0)}} 
 + \lambda\, \frac{\D\sigma^{(0)}\,\D\sigma_{L-R}^{(1,1)} - 
   \D\sigma^{(1)} \,\D\sigma_{L-R}^{(0,1)}}{2\,\big(\D\sigma^{(0)}\big)^2} \, .
\end{align}
\end{subequations}
The difference between the expanded and unexpanded versions of $A_{LR}^{\text{NLO}}$ is of order $\alpha\lambda$ or has terms $\lambda^2$ with more than one insertion of a higher-dimensional operator.

\begin{figure}
\centering  
    \includegraphics[height=2.4cm]{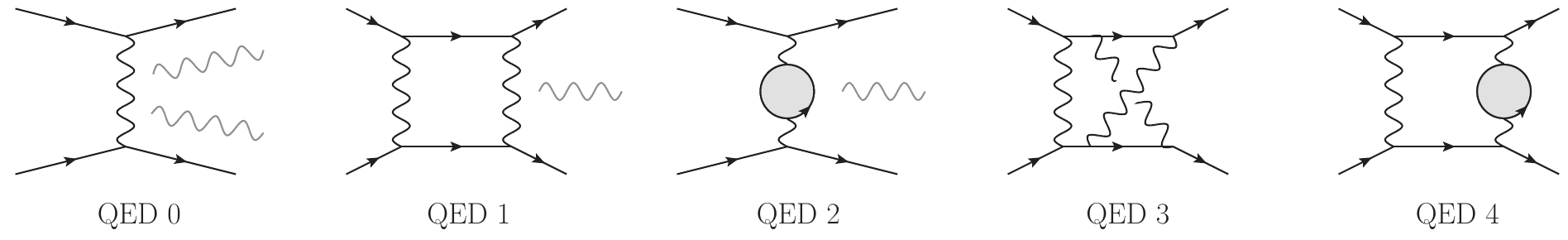} \\[20pt]
    \includegraphics[height=2.4cm]{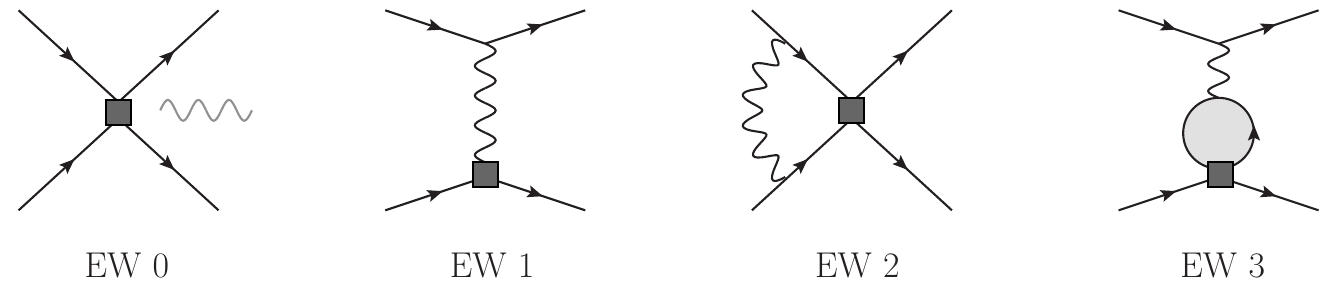}
\\[10pt]
    \caption{Representative diagrams for the amplitude~\eqref{AmpMoller}. The top line are LO, NLO, and NNLO QED diagrams. The bottom line shows LO and NLO diagrams with a single insertion of a higher-dimensional operator. The required real amplitudes are indicated by additional (grey) external photons.  }
    \label{fig:diag}
\end{figure}

To compute the required cross sections we need the  M{\o}ller scattering amplitude
\begin{subequations}\label{AmpMollerALL}
\begin{align}\label{AmpMoller}
  \cA_n&= \alpha\,\cA_{n}^{(0)} 
  + \lambda\,\cA_{n}^{(0,1)} 
  + \alpha^2\,\cA_{n}^{(1)} 
  + \alpha \lambda\,\cA_{n}^{(1,1)} 
  + \alpha^3\,\cA_{n}^{(2)} + 
  \ldots\  \, ,
\end{align}
where we indicate corrections beyond our approximation by the ellipses. Typical diagrams are shown in Figure~\ref{fig:diag}.  The subscript $n=2$ indicates two
particles in the final state. For the real corrections, we also need the amplitudes
with up to two additional photons (depicted in grey in Figure~\ref{fig:diag}) in the final state
\begin{align}\label{AmpMollerR}
  \cA_{n+1}&= \alpha^{3/2}\,\cA_{n+1}^{(0)} 
  + \alpha^{1/2} \lambda\,\cA_{n+1}^{(0,1)} 
  + \alpha^{5/2}\,\cA_{n+1}^{(1)} 
  + \ldots\  \, , \\ \label{AmpMollerRR}
  \cA_{n+2}&= \alpha^2 \cA_{n+2}^{(0)} + \ldots \, .
\end{align}
\end{subequations}
For the pure QED contribution at LO, we need $\cA_n^{(0)}$ and integrate the squared matrix element $\cM_{n}^{(0)} = |\cA_n^{(0)}|^2$ over the phase space. For virtual and real corrections at NLO we need $\cA_n^{(1)}$ and $\cA_{n+1}^{(0)}$, while $\cA_n^{(2)}$, $\cA_{n+1}^{(1)}$, and $\cA_{n+2}^{(0)}$ are required at NNLO. This includes vacuum polarisation (VP) contributions, indicated by grey blobs in QED~2 and QED~4 of Figure~\ref{fig:diag}. The leptonic part of the VP can be computed in perturbation theory. The hadronic part will be discussed in Section~\ref{sec:dirtyWorld}.  A complete calculation of M{\o}ller scattering at NNLO in QED has been presented in~\cite{Banerjee:2021qvi}. 

If we are interested in parity-violating effects, the leading
contributions to the squared matrix elements are 
\begin{subequations}\label{AmpSqMoller}
\begin{align} \label{AmpSqMollerLO}
  \cM_n&= \alpha \lambda\,\cM_n^{(0,1)} 
  + \alpha^2 \lambda\,\cM_n^{(1,1)} + \ldots \,, \\
  \cM_n^{(0,1)}&= 2\,\Re\big(\cA_n^{(0)\,*} \cA_n^{(0,1)} \big)\,, \\
  \cM_n^{(1,1)}&= 2\,\Re\big(\cA_n^{(1)\,*} \cA_n^{(0,1)} \big)
  + 2\,\Re\big(\cA_n^{(0)\,*} \cA_n^{(1,1)} \big) \, ,
\end{align}
\end{subequations}
where the ellipses stand for higher-order terms as well as parity-conserving pure QED contributions $\cM_n^{(0)}$, $\cM_n^{(1)}$, and $\cM_n^{(2)}$.
We will refer to $\cM_n^{(0,1)}$ and $\cM_n^{(1,1)}$ as the LO and
virtual NLO (parity-violating) matrix elements, respectively. Of course, $\cM_n^{(1,1)}$
contains IR singularities that will be cancelled by the corresponding
real corrections. Regardless of the appearance of IR singularities,
the real corrections have to be added in order to obtain
physical results. Thus, we also need the real matrix element
\begin{align} \label{AmpSqRealMoller}
  \alpha^2 \lambda\, \cM_{n+1}^{(0,1)}&
  =\alpha^2 \lambda \, 2\,\Re\big(\cA_{n+1}^{(0)\,*} \cA_{n+1}^{(0,1)} \big)  \, .
\end{align}

These diagrams with $\mathcal{O}_i$ insertions are illustrated in the bottom panel of Figure~\ref{fig:diag}.  At tree level, $\cA_{n}^{(0,1)}$, only four-fermion operators contribute, with an additional photon for $\cA_{n+1}^{(0,1)}$, required for the real NLO corrections. The counting of the dipole operator appearing in EW~1 of Figure~\ref{fig:diag} includes a factor $\sqrt{\alpha}$ and we treat it in the same way as dimension 6 operators. The reason is that in the SM this operator is loop induced and the corresponding Wilson coefficient always contains a factor $m_{\ell}$.
Hence, it contributes to $\cA_n^{(1,1)}$. Also the loop diagrams, either with a photon exchange or fermion loop contribute to $\cA_n^{(1,1)}$. As for the VP contributions in QED, the penguin diagrams EW~3 can be computed perturbatively for lepton loops but lead to difficulties for hadronic loops at scales $\mu_s \lesssim 2\,\GeV$.

We can treat \eqref{eq:Lagb} as a fundamental Lagrangian with externally given Wilson coefficients $C_i(\mu_s)$. The Wilson coefficients then have no connection to the parameters of $\cL_\text{QED}$ and we are free to use whatever renormalisation scheme we want for $\cL_\text{QED}$. Typically, for low-energy processes the on-shell scheme is used for the coupling $\alpha$ and the masses $m_\ell$. The Wilson coefficients themselves also need a renormalisation scheme and the canonical choice is $\MS$, apart from finite remnants of evanescent operators, as discussed in~\cite{Dekens:2019ept}.

If we want to make contact with the SM, the Wilson coefficients $C_i(\mu_s,\alpha,m_\ell,M)$ are determined by matching LEFT to the SM at the high scale $\mu_h\sim \Mew$ and using the RGE to evaluate them at the low scale $\mu_s$. This allows to resum logarithms of the form $\ln(\mu_h/\mu_s)$. 
In this case, the Wilson coefficients also depend on the parameters $\alpha$ and $m_\ell$ of $\cL_\text{QED}$, and many others generically denoted by $M$. In order to resum all logarithms, we have to use the $\MS$ coupling $\bar{\alpha}(\mu)$ and masses $\bar{m}_\ell(\mu)$. We can then obtain the RGE-improved Wilson coefficients at the low scale but expressed in terms of the $\MS$ renormalised parameters of $\cL_\text{QED}$. In order to combine this with fixed-order QED computations which are typically carried out in the on-shell scheme, we need to adjust the schemes. For our results for the MOLLER experiment in Section~\ref{sec:ResMoller}, we choose to convert the fixed-order results to the on-shell scheme but keep the Wilson coefficient expressed in terms of $\MS$ quantities. Since the conversion is done at a low scale, this can be done without generating large logarithms. This procedure ensures that the obtained results re-expanded to a fixed order in $\alpha$ agree with computations in the SM to the same order. It also guarantees the inclusion of all large logarithms at the order determined by the order of the matching and the order of the anomalous dimensions used in the running. In our current case this is NLO and LL, and implies that the matching is done at one loop, at least for those operators that appear in the tree-level result. However, the LEFT setup allows for a systematic improvement of the perturbative approximation. In particular, the availability of the two-loop anomalous dimensions allows for a resummation at the NLL level and we will investigate the effect of these contributions. 

To illustrate our computation in the LEFT framework, we present explicit results for $A_{LR}^{\text{vNLO exp}}$, the virtual part of the NLO asymmetry~\eqref{ALRexp}, strictly expanded in $\alpha$.
In this case, the IR singularities cancel between the two terms of $\cO(\lambda)$, even if only the virtual parts are included. Indeed, the IR singularities factorise as $\D\sigma_{L-R}^{(1,1)} \rightsquigarrow \cE\,\D\sigma_{L-R}^{(0,1)}$ and $\D\sigma^{(1)} \rightsquigarrow \cE\,\D\sigma^{(0)}$, where $\cE$ is the same IR divergent integrated eikonal factor. While this does not imply that the real corrections should be neglected, the partial result 
\begin{subequations} \label{eq:ALRexplicit}
\begin{align} \label{eq:ALRtree}
    A_{LR}^{\text{LO exp}}&=
    \frac{2s}{e^2\,Q_e^2}    
%    \revi{\frac{s}{2\pi\bar\alpha\,Q_e^2}}
    \,\bigl(C_{ee,1111}^{V,LL}-C_{ee,1111}^{V,RR}\bigr)\,F_{00}(y) \,, \\
	A_{LR}^{\text{vNLO exp}}&=\frac{s}{8\pi^2}\,
    \biggl\{\bigl(C_{ee,1111}^{V,LL}-C_{ee,1111}^{V,RR}\bigr)\,\biggl[F^{\Box}(y)+F_{03}^{\triangle}(y)+F_{20}(y,0)+2F_{03}(y,0) 
    \nonumber \\[-4pt]
	&\hspace{143.6pt}+\hspace{-7.4pt}\sum\limits_{p\,\in\{2,3\}}\hspace{-7.4pt}F_{20}(y,m_{e,p})+N_c\hspace{-7.5pt}\sum\limits_{q\,\in\{u,d\}}\hspace{-2pt}\frac{Q_q^2}{Q_e^2}\sum\limits_pF_{20}(y,m_{q,p})\biggr] 
    \nonumber \\[-2pt]
	&\hspace{-1pt}+\hspace{-7.4pt}\sum\limits_{p\,\in\{2,3\}}
    \biggl[\Bigl(4C_{ee,11pp}^{V,LL}+C_{ee,11pp}^{V,LR}-C_{ee,pp11}^{V,LR}-4C_{ee,11pp}^{V,RR}\Bigr)\,F_{03}(y,m_{e,p}) \nonumber \\[-6pt]    
    &\hspace{39.1pt}+\frac{4}{3}\,\bigl(C_{ee,11pp}^{V,LL}-C_{ee,11pp}^{V,RR}\bigr)\,F_{00}(y)\biggr] 
    \nonumber \\
	&\hspace{-1pt}+N_c\hspace{-7.5pt}\sum\limits_{q\,\in\{u,d\}}\hspace{-2pt}
    \frac{Q_q}{Q_e}\sum\limits_p
    \Bigl(C_{eq,11pp}^{V,LL}+C_{eq,11pp}^{V,LR}-C_{qe,pp11}^{V,LR}-C_{eq,11pp}^{V,RR}\Bigr)\,F_{03}(y,m_{q,p})\biggr\} \, , \label{eq:ALRvirt}
\end{align}
\end{subequations}
with $e\equiv \sqrt{4\pi\bar\alpha}$, allows for a comparison with results presented in the literature. The functions $F_i$ are listed in Appendix~\ref{sec:appendixexplicit} and depend on $y$ and masses. For brevity, we present the analytic results in the limit $m_e\to{0}$, but we keep mass effects of the quarks, the muon, and the tau. Hence, \eqref{eq:ALRtree} is equivalent to \eqref{ALRLO}. The results with full $m_e$ dependence are available in electronic form and can be obtained 
from the public repository
\begin{quote}
    \url{https://gitlab.com/mule-tools/left-tools}
\end{quote}

The terms coming from the interference of diagrams [QED~1]$\otimes$[EW~0] and [QED~0]$\otimes$[EW~2] of Figure~\ref{fig:diag} (without external photons) are collected in $F^{\Box}(y) = F_{10}(y) + F_{02}(y)$, which reproduces effects of SM one-loop box diagrams. The terms with the dipole operator [QED~0]$\otimes$[EW~1] cancel between $\D\sigma_L$ and $\D\sigma_R$ for real Wilson coefficients (i.e. for vanishing electric dipole moment). 
The terms proportional to $F_{20}(y,m_i)$ come from the interference of [QED~2]$\otimes$[EW~0], where the sum is over all active fermion species other than the electron (the top is integrated out). The contributions from closed electron loops are included in $F_{20}(y,0)$. 
Finally, the $F_{03}$ terms come from the interference of [QED~0]$\otimes$[EW~3] with a fermion loop. Again, the results for the electron proportional to $F_{03}(y,0)$ is related to the one of massive fermions $F_{03}(y,m)$. The factor 2 as well as the additional terms in the third line of \eqref{eq:ALRvirt} are explained in the Appendix~\ref{sec:appendixexplicit}. In the case of an electron in the loop, there is also a [QED~0]$\otimes$[EW~3] contribution with an open fermion line, resulting in $F_{03}^{\triangle}$. 

The result \eqref{eq:ALRvirt} is given  for $\mu_s > m_b$ and initially in terms of $\bar\alpha$, the $\MS$ coupling within LEFT. In order to obtain an expression for $\mu_s < m_b$ the $b$ quark has to be integrated out. This results in threshold corrections
\begin{align} \label{eq:thresholdcorr}
  C_{ee,1111,f=7}^{V,LL/RR}(m_b) &=  
   C_{ee,1111,f=8}^{V,LL/RR}(m_b) -\bar\alpha_{f=7}^2\,Q_e^2Q_d^2\,\frac{2N_c}{15\,m_b^2}\,.
\end{align}
Furthermore, the terms with the $b$ quark ($q=d, p=3$) do not contribute any longer for $\mu_s < m_b$. This procedure has to be repeated at every threshold and ensures that heavy fermion contributions included in $F_{20}$ properly decouple in the limit $m_i\to\infty$. In practice, if we stop the perturbative running at $\mu\simeq{2}\,\GeV$ only the $b$ threshold is considered. At the low scale it is possible to convert to the on-shell scheme for $\alpha$. The change induced in $F_{20}$ through this renormalisation-scheme conversion is given explicitly in Appendix~\ref{sec:appendixexplicit}. 

In~\eqref{eq:ALRexplicit}, only those operators of LEFT that are induced by the SM are visible.
In principle, there are contributions from scalar operators, but they vanish in the limit $m_e\to 0$. 
They are also included in the repository linked above.
The fact that the contribution to $A_{LR}$ of LEFT operators not induced through the SM is suppressed by the small electron mass reduces the sensitivity of $A_{LR}$ on new physics.
Of course, there is still the possibility that new physics is embedded in SM-like Wilson coefficients in \eqref{eq:ALRexplicit}.
Typical examples of BSM models mentioned in connection with MOLLER are a dark $Z$~\cite{Davoudiasl:2012ag} or doubly charged scalar~\cite{Dev:2018sel}, which both contribute to vector operators. On the other hand, leptoquarks are an example of BSM physics, leading to scalar operators~\cite{Bischer:2021jqn}. At the high scale, these are lepton-quark operators that mix into pure lepton operators.

As a check, we have verified that the EFT result \eqref{eq:ALRexplicit} agrees with an explicit computation within the full SM after expanding the latter in $\mu_s/\Mew$. This requires the one-loop and tree-level matching coefficients in \eqref{eq:ALRtree} and \eqref{eq:ALRvirt} respectively. The advantage of \eqref{eq:ALRexplicit}  compared to the SM calculation is that inserting the LL RGE-improved Wilson coefficients resums all logarithms of the form $(\alpha L)^n$.
In addition, we have checked our results against the first calculation of the one-loop corrections to $A_{LR}$~\cite{Czarnecki:1995fw}. In that work, results are presented in a renormalisation scheme with 
$\alpha$ and $M_Z$ in the on-shell scheme, $s_W$ in the $\overline{\rm MS}$ scheme and $M_W$ in a scheme such that the relation $c_W=M_W/M_Z$ is preserved to all orders in perturbation theory. 
Moreover, the $G_\mu$ scheme was used as input parameter scheme.
As described e.g. in~\cite{Denner:2019vbn}, in such a scheme the 
result is expressed in terms of the Fermi constant $G_\mu$ extracted from the muon decay.
After changing the renormalisation and input parameter scheme used in this work to the 
ones used in \cite{Czarnecki:1995fw}, we find  that both results are in agreement. 
In particular, this requires to adapt the renormalisation scheme in the matching results.

For the calculation, diagrams were generated with {\tt FeynArts}~\cite{Hahn:2000kx}, while the Dirac algebra was done using {\tt Package-X}~\cite{Patel:2015tea,Patel:2016fam} and {\tt Tracer}~\cite{Jamin:1991dp}.
The one-loop integration was done by using {\tt Package-X} and {\tt PVReduce}~\cite{Patel:2015tea}.
In order to check the calculation in the EFT against a calculation in the full SM, we have expanded the scalar 
integrals appearing in the SM by using the method of regions.
We have checked the expansion of the scalar integrals numerically, also using {\tt Collier}~\cite{Denner:2016kdg} in combination with {\tt CollierLink}~\cite{Patel:2016fam}.

\subsection{Including nonperturbative hadronic contributions} \label{sec:dirtyWorld}
Hadronic contributions enter the one-loop calculation in LEFT through quark loops in the photon two-point function and penguin diagrams corresponding to the $\gamma Z$ two-point function in the SM.
Such contributions are typically expressed in terms of current-current correlators $\Pi$ that are connected to the self-energy as follows
\begin{align}
    \Sigma(q^2) = \Sigma(0) + q^2 \, \Pi(q^2)\,.
\end{align}
Relevant for this work are the
charge-charge $\Pi_{\gamma\gamma}$ and the charge-isospin $\Pi_{\gamma 3}$ contributions, where
$\Pi_{\gamma 3}$ is related to the third component of the isospin current of the $Z$ coupling which leads to
\begin{align}
\label{eq:gammaZpi}
    \Pi_{\gamma Z}(q^2) = \frac{1}{c_W s_W} \left(\Pi_{\gamma 3}(q^2) - s_W^2 \Pi_{\gamma \gamma}(q^2) \right)\,.
\end{align}
The isospin-isospin $\Pi_{3 3}$ do not enter at dimension 6. 
They are part of $\Pi_{Z Z}$.
In~\eqref{eq:gammaZpi}, $s_W$ is renormalised in the on-shell scheme. Since we are only considering EW effects at NLO, we will later use $\hat{s}_W$ as the numerical value in~\eqref{eq:gammaZpi} and the subsequent equations.

While the combination $\Pi_{\gamma\gamma}(q^2)-\Pi_{\gamma\gamma}(0)$ appears in standard QED calculations and is therefore well established and can be obtained from different codes~\cite{Hagiwara:2006jt,Ignatov:hvp,Jegerlehner:hvp19} (see \cite{Aliberti:2024fpq} for an overview of codes used by different Monte Carlo tools), the evaluation of $\Pi_{\gamma Z}(q^2)$ is a recurring topic in the literature~\cite{Tomalak:2025tls,Du:2019evk,Erler:2022ckm,Erler:2017knj,Marciano:1982mm,Marciano:1993jd,Czarnecki:1995fw,Hoferichter:2025yih}.
In this work, we follow a simplified approach to model all hadronic contributions consistently by using the Fortran library {\tt alphaQED}~\cite{Jegerlehner:2001ca, Jegerlehner:2006ju, Jegerlehner:2011mw, Jegerlehner:hvp19}.

{\tt alphaQED} provides the quantities
\begin{subequations}
\label{eq:deltaalphas}
\begin{align}
    \Delta\alpha_{\rm had}(q^2) &= - \left(\Real \Pi_{\gamma\gamma,{\rm had}}(q^2)-\Pi_{\gamma\gamma,{\rm had}}(0)\right)\,, \\
    \Delta\alpha_{2, {\rm had}}(q^2) &= - \left(\Real \Pi_{\gamma 3,{\rm had}}(q^2)-\Pi_{\gamma 3,{\rm had}}(0)\right)\,,
\end{align}
\end{subequations}
by extracting them from $e^+ e^-\to\text{hadrons}$ data.
While {\tt alphaQED} is the only code available for the evaluation of $\Delta\alpha_{2, {\rm had}}(q^2)$, its results have been verified against lattice QCD~\cite{Burger:2015lqa}.
From~\eqref{eq:deltaalphas} it follows that the nonperturbative contributions to the difference $\Pi_{\gamma Z, {\rm had}}(q^2)-\Pi_{\gamma Z, {\rm had}}(0)$ can be expressed through quantities given by {\tt alphaQED}
\begin{align}
    \left(\Real \Pi_{\gamma Z,{\rm had}}(q^2)-\Pi_{\gamma Z,{\rm had}}(0)\right) &=  -\frac{1}{c_W s_W}\left(\Delta\alpha_{2,{\rm had}}(q^2)-\Delta\alpha_{\rm had}(q^2)s_W^2\right)\,.
    \label{eq:relgZtoalphaQED}
\end{align}
However, the evaluation of just $\Pi_{\gamma Z, {\rm had}}(q^2)$ individually, as required in EW calculations, demands another preferably perturbative input.
For the (one-loop) self energy we can write
\begin{align}
    \Pi^{(1)}_{{\rm had}}(q^2,\mu) =& 
    \left(\Pi_{{\rm had}}(q^2)-\Pi_{{\rm had}}(0)\right) - 
    \left(\Pi_{{\rm had}}(M_Z^2)-\Pi_{{\rm had}}(0)\right) \nonumber
    \\ &+ \Pi^{(1)}_{{\rm had}}(M_Z^2,\mu)
    \label{eq:relatedgZtoMZ}\,.
\end{align}
The second line of~\eqref{eq:relatedgZtoMZ}, the self energy evaluated at a heavy scale, here chosen to be $M_Z$, has two advantages: it can be evaluated perturbatively with vanishing quark masses, and it preserves the $\mu$ dependence of the l.h.s. of~\eqref{eq:relatedgZtoMZ}.
At one loop, we obtain in perturbation theory
\begin{subequations}
\begin{align}
    \Pi^{(1),\MS}_{\gamma Z,{\rm had}}(M_Z^2,\mu) &= - \frac{\alpha}{\pi} \frac{44 s_W^2 - 21}{324 c_W s_W} \left(5+3\ln\left(-\frac{\mu^2}{M_Z^2}\right)\right)N_c\,, \\
    \Pi^{(1),\MS}_{\gamma \gamma,{\rm had}}(M_Z^2,\mu) &= - \frac{\alpha}{\pi} \frac{11}{81} \left(5+3\ln\left(-\frac{\mu^2}{M_Z^2}\right)\right) N_c\,.
\end{align}
\end{subequations}
Hence, in~\eqref{eq:relatedgZtoMZ} we have related the nonperturbative ($\MS$ subtracted) $\Pi^{(1)}_{{\rm had}}(q^2,\mu)$ at a low scale to two nonperturbative quantities provided by {\tt alphaQED} (first line of~\eqref{eq:relatedgZtoMZ}) and to the perturbative ($\MS$ subtracted) $\Pi^{(1)}_{{\rm had}}(M_Z^2,\mu)$, preserving their $\mu$ dependencies.
With this prescription, we do not only evaluate $\Pi_{\gamma\gamma,\rm had}(q^2)-\Pi_{\gamma\gamma,\rm had}(0)$ from the on-shell renormalised QED part but also $\Pi_{\gamma Z, {\rm had}}(q^2)$ by using the most recent version of {\tt alphaQED}, {\tt alphaQEDc23}.
An implementation of more accurate models for $\Pi_{\gamma Z}$ is planned for the future.
A first comparison to~\cite{Hoferichter:2025yih} shows that the values of~\eqref{eq:relgZtoalphaQED} at $q^2 = - M^2_Z$ are consistent within uncertainties.
At momentum transfers in the range of the MOLLER experiment, differences at the percent level can be anticipated.

In the nonperturbative regime, the quark fields in LEFT are not the adequate degrees of freedom to describe hadronic effects in the dimension 6 part of the electron-photon three-point function,
\begin{equation}
    i\Gamma_{\text{had}}^{eeA}(p,k,q)=\hspace{-3pt}
    \begin{tikzpicture}[scale=0.6,baseline=-\the\dimexpr\fontdimen22\textfont2\relax]
	\draw[photon1] (0.2,0) -- (1.8,0);
	\draw[fermion1] (0,1) -- (0.2,0);
	\draw[fermion2] (0.2,0) -- (0,-1);
	\fill[gray!30] (0.6,0) circle[radius=0.4];
	\draw[
	postaction={decorate},
	decoration={markings, mark=at position 0.75 with {\draw[-{Latex[length=3pt]}] (-1pt,0) -- (1pt,0);}}
	] (0.6,0) circle[radius=0.4];
	\filldraw[black!80,draw=black!80] (-0.12+0.2,-0.12) rectangle (0.12+0.2,0.12);
    \draw[-{Latex[length=1mm, width=0.9mm]}] (-0.16,-0.8) -- (-0.05,-0.3) node[midway, left] {\scriptsize$p$};
	\draw[-{Latex[length=1mm, width=0.9mm]}] (-0.16,0.8) -- (-0.05,0.3) node[midway, left] {\scriptsize$k$};
	\draw[-{Latex[length=1mm, width=0.9mm]}] (1.6,0.27) -- (1.1,0.27) node[midway, above] {\scriptsize$q$};
\end{tikzpicture}\;\;.
\end{equation}
As a consequence, the effects due to $\gamma Z$ mixing in the SM cannot be expressed in terms of the Wilson coefficients in the given basis. 
Instead, they have to be related to data through $\Pi_{\gamma Z, {\rm had}}(q^2)$. This implies the Lorentz decomposition
\begin{equation}\label{eq:leptonZvertexdecomp}
    \Gamma_{\text{had}}^{eeA}(p,k,q)=\biggl(g_{\mu\nu}-\frac{q_{\mu}q_{\nu}}{q^2}\biggr)\gamma^{\nu}P_L\,\Gamma^{V,L}(q^2)+q_{\mu}P_L\,\Gamma^{S,L}(q^2)+\sigma_{\mu \nu } q^\nu P_L\,\Gamma^{T,L}(q^2)+\{L\leftrightarrow R\}
\end{equation}
at one loop. 
The ($\overline{\text{MS}}$ subtracted) finite parts of the functions $\Gamma^i(q^2)$ are matched by equating~\eqref{eq:leptonZvertexdecomp} to the one-loop expression for the hadronic contributions to the electron-photon vertex in the SM, written in terms of $\Pi_{\gamma Z,\text{had}}^{\overline{\text{MS}}}(q^2)$. In fact, among the functions entering lepton-lepton scattering in LEFT, only $\Gamma^{V,L/R}$ receive a non-zero contribution in the SM. After subtracting the perturbative pole in LEFT, we obtain
\begin{subequations}
\begin{align}
    \Gamma_{\overline{\text{MS}}}^{V,L,(1)}(q^2)&=\frac{\sqrt{4\pi\alpha}\,(1+2 s_W^2\,Q_e)}{2 c_W s_W M_Z^2}\,q^2\,\Pi_{\gamma Z,\text{had}}^{\overline{\text{MS}}}(q^2)\,,\\
    \Gamma_{\overline{\text{MS}}}^{V,R,(1)}(q^2)&=\frac{\sqrt{4\pi\alpha}\,s_W\,Q_e}{c_W\,M_Z^2}\,q^2\,\Pi_{\gamma Z,\text{had}}^{\overline{\text{MS}}}(q^2)\,.
\end{align}
\end{subequations}

The explicit results for all $\Gamma^i$ and details on our use of {\tt alphaQED} can be found in the public repository that is linked in Section~\ref{sec:perturbative}.
The functions in~\eqref{eq:leptonZvertexdecomp} depend in particular on $\alpha$, $s_W$ and $M_Z$.
Within our perturbative expansion in LEFT we are free to choose the scale $\mu$ at which they are evaluated.
When using those functions in Section~\ref{sec:ResMoller} in \mcmule{}, we evaluate $s_W$ and $M_Z$ at $\mu=\Mew$ given in~\eqref{eq:inputval} in Appendix~\ref{sec:appendixinput} and take the on-shell value for $\alpha$.

To allow for any potential BSM application, a use in the perturbative regime or for modelling the $\gamma Z$ exchange with quark masses, \mcmule{} also allows to switch to a version where~\eqref{eq:leptonZvertexdecomp} is expressed in terms of Wilson coefficients involving leptons and light quarks.

\section{Results} \label{sec:results}
In what follows, we present two sets of results.
In Section~\ref{sec:ResPerturbative} we illustrate how the virtual corrections change under different RGE improvements.
To fully illustrate LEFT, we restrict us in this section to a centre-of-mass energy where the light quarks can be described within perturbation theory.
In Section~\ref{sec:ResMoller}, we present the full results tailored to the MOLLER experiment, where we only apply LL RGE-improved Wilson coefficients and connect it with NNLO in QED as well as nonperturbative hadronic effects. 
However, implementing RGE improvements beyond LL into \mcmule{} is straightforward.

\subsection{Perturbative virtual results} \label{sec:ResPerturbative}
We start by presenting results for M{\o}ller scattering at $\sqrt{s}=2\,\GeV$, using a purely perturbative approach. Furthermore, in this subsection we restrict ourselves to (RGE-improved) virtual corrections of the expanded $A_{LR}^\text{exp}$ as defined in \eqref{ALRexp}, and neglect real corrections. While such results are unphysical and not directly connected to any experiment, they still illustrate several important points. The results in this subsection are expressed in terms of the $\MS$ coupling $\bar\alpha(\mu_s)$ and are obtained for incoming electrons with 100\% longitudinal polarisation.

\begin{figure}[t]
	\centering 
	\includegraphics[width=0.8\textwidth]{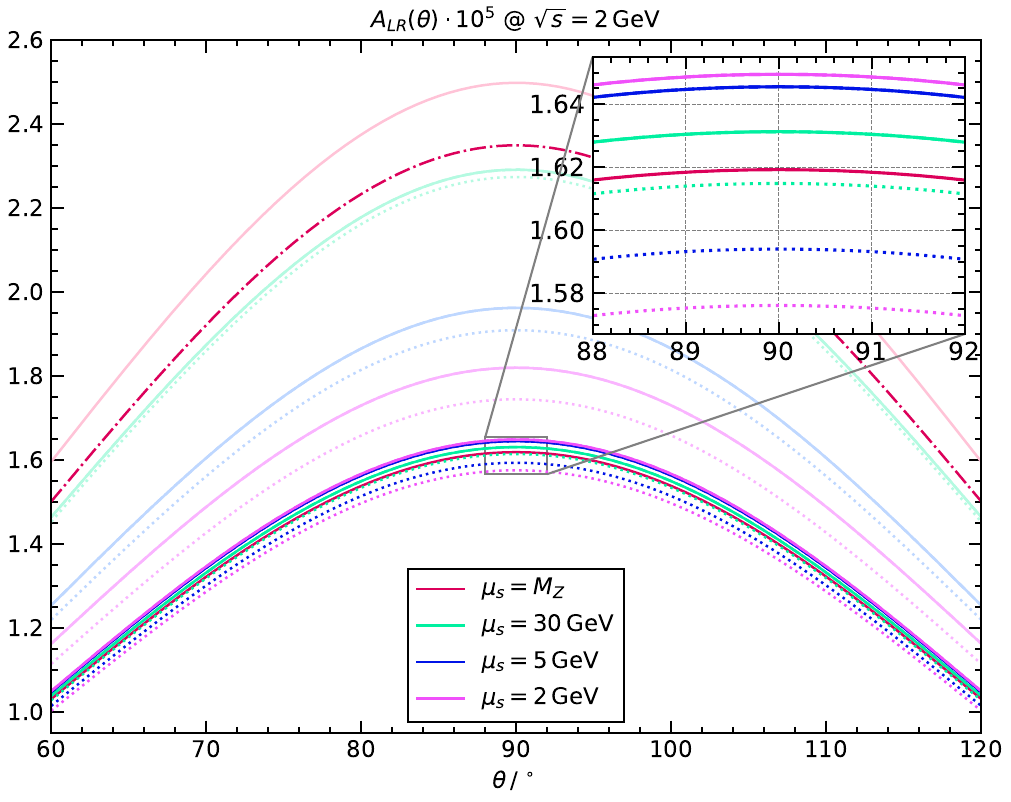}
    \caption{\label{fig:ALRvirtmu} The virtual part of  $A^{\text{exp}}_{LR}$ for M{\o}ller scattering at $\sqrt{s}=2\,\text{GeV}$ at LO (light colours) and NLO (dark colours) in RGE-improved perturbation theory at LL (solid lines) and NLL (dotted lines) accuracy for different values of $\mu_s$ from the EW scale down to $\sqrt{s}$. 
    For $\mu_s=2\,\text{GeV}$, threshold corrections are properly taken into account in LEFT by integrating out the $b$ quark at its mass threshold, up to one-loop order and dimension six.
    }
\end{figure}

In Figure~\ref{fig:ALRvirtmu} we show $A_{LR}^{\text{LO exp}}$ and $A_{LR}^{\text{vNLO exp}}$ for various values of $\mu_s$ in order to illustrate the effects due to the resummation of large logarithms. The results with $\mu_s=M_Z$, depicted as red lines, correspond to using the Wilson coefficients at the matching scale and, hence, to strict fixed-order results without resummation. As is well known, this leads to extremely large corrections from LO (light colours) to NLO (dark colours), substantially reducing the asymmetry. Lowering the scale $\mu_s$ towards the natural value $\sqrt{s}=2\,\GeV$ leads to significantly smaller corrections. To be concrete, this means we use the LL RGE-improved Wilson coefficients at $\mu_s$ and evaluate \eqref{eq:ALRtree} and \eqref{eq:ALRvirt}. In doing so, for $\mu_s=\sqrt{s}$ we resum all leading logarithms of the form $\ln(\sqrt{s}/\Mew)$. The LL RGE-improved LO and NLO results for $\mu_s\in\{30\,\GeV, 5\,\GeV, 2\,\GeV\}$ are shown in green, blue, and magenta. For the preferred choice $\mu_s=\sqrt{s}=2\,\GeV$ the corrections amount to roughly 15\%. Thus, in RGE-improved perturbation theory, the corrections are under much better control. The still substantial size of 15\% is connected to the accidental suppression factor $1-4 s_W^2$ of the LO result. Furthermore, the LL NLO results show a remarkable stability under $\mu_s$ variations. In fact, the last step in the RGE, from $5\,\GeV$ to $2\,\GeV$ only results in a change of less than half a percent. For this step, it is necessary to include the threshold corrections that appear when integrating out the $b$ quark. 

The red dashed-dotted line in Figure~\ref{fig:ALRvirtmu} is the NLO result at $\mu_s=M_Z$ excluding quark loops. Comparing this to the LO and full NLO result confirms that the bulk of the NLO corrections are due to quark loops. This fact motivates the extensive effort in the literature to improve the theoretical description of hadronic contributions to $A_{LR}$.

In order to illustrate the potential impact of logarithms at NLL accuracy, we use a version of {\tt DsixTools}~\cite{Celis:2017hod,Fuentes-Martin:2020zaz} with the two-loop anomalous dimensions~\cite{Aebischer:2025hsx} implemented. In addition, we include the two-loop beta function in the evolution of $\bar\alpha(\mu)$ and additional running effects in QCD beyond NLL accuracy. The numerical impact of the latter is negligible. The LO (NLO) results with these NLL RGE-improved coefficients are depicted as light (dark) dotted lines. As can be seen in the inset of Figure~\ref{fig:ALRvirtmu}, switching from LL Wilson coefficients to NLL in the NLO result has a notable impact, in particular for $\mu_s=\sqrt{s}=2\,\GeV$. It is also evident that the LL NLO scale-variation band does not seem to be a reliable indicator for a theoretical error. From these observations it appears that a theoretical description of $A_{LR}$ for very low scales at the percent level requires the inclusion of corrections beyond LL NLO. 

\subsection{Results for the MOLLER experiment} \label{sec:ResMoller}

In order to obtain results that are directly applicable to the MOLLER experiment, we have implemented the LL RGE-improved NLO calculation within LEFT in the public code \mcmule{}\cite{Banerjee:2020rww}. The LO and NLO EW corrections $\D\sigma_{L/R}^{(0,1)}$ and $\D\sigma_{L/R}^{(1,1)}$ are evaluated with LL RGE improved Wilson coefficients $C_i(\mu_s)$.
As mentioned in Section~\ref{sec:framework}, to combine this with the already available NNLO QED calculation~\cite{Banerjee:2021qvi} in the on-shell scheme we also express \eqref{eq:ALRexplicit} in terms of the on-shell coupling~$\alpha$.

The implementation of these results in \mcmule{} allows us to calculate any infrared-safe observable in a fully differential way, including arbitrary experimental cuts. We will present results in the centre-of-mass frame and in the laboratory frame.
All raw data, analysis pipelines, and plots can be found at~\cite{McMule2022:library}
\begin{quote}
    \url{https://mule-tools.gitlab.io/user-library/moller-scattering/moller}
\end{quote}

In order to mimic the MOLLER detector we follow their technical design report~\cite{MOLLERTDR} and use 
\begin{subequations} \label{eq:cuts}
\begin{align}
    &\mbox{Longitudinal polarisation } P=90\%  \label{cutsPol} \\
    &E_{\rm beam} = 11\,\GeV \quad (\sqrt{s}=106.031 \,\MeV) \label{cutsEnergy} \\ 
    &50^\circ \le \theta_{3,4} \le 130^\circ \quad
    \mbox{for electron scattering angles in the centre-of-mass frame} \label{cutsThetaCoM} \\
    &5\,\mrad \le \tilde{\theta}_{3,4} \le 20\,\mrad \quad
    \mbox{for electron scattering angles in the laboratory frame} \label{cutsThetaLab}
\end{align}
\end{subequations}
It is understood that both electrons have to satisfy the angular cuts in order for the event to be selected. 
For the results in the laboratory frame, we apply both angular cuts,~\eqref{cutsThetaLab} and~\eqref{cutsThetaCoM}.
Furthermore, we are completely inclusive with respect to additional photon radiation. 

As an example for an observable in the centre-of-mass frame, we define
\begin{align} \label{eq:CoMtheta}
    \theta &= \frac{1}{2} \left(\pi - \vert \theta_3 - \theta_4 \vert\right) \, .
\end{align}

This observable takes into account that the two scattered electrons are indistinguishable. For tree-level kinematics, $\theta$ corresponds to the scattering angle in the range $50^\circ\le\theta_{3,4}\le 90^\circ$.

\begin{figure}[t]
\centering  
    \includegraphics[width=0.9\textwidth]{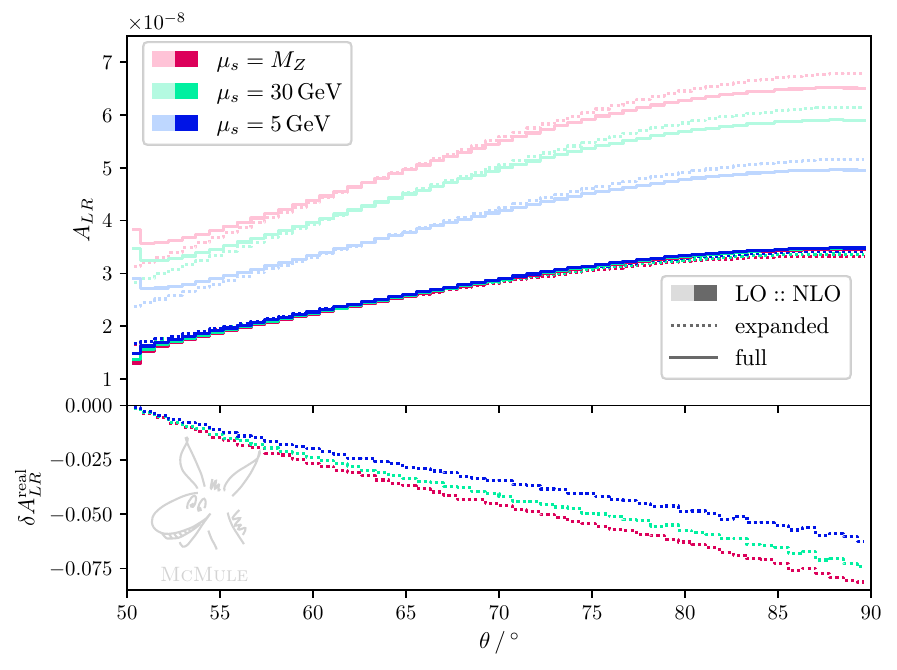}
\\[10pt]
    \caption{$A_{LR}$ for the MOLLER experiment, as a function of $\theta$ defined in~\eqref{eq:CoMtheta} in the centre-of-mass frame at LO and NLO in RGE-improved perturbation theory, for three choices of $\mu_s$. Solid and dotted lines correspond to \eqref{ALRfull} and \eqref{ALRexp} respectively. The bottom panel shows the impact of real corrections for the expanded version of $A_{LR}$.}
    \label{fig:mollercm}
\end{figure}

The results are shown in Figure~\ref{fig:mollercm} for three choices of $\mu_s$. The solid lines correspond to the full $A_{LR}$ \eqref{ALRfull} at LO (light colours) and NLO (dark colours) in LL RGE-improved perturbation theory. The dotted lines show the corresponding expanded results \eqref{ALRexp}. There are noticeable differences between the two at LO, whereas at NLO the full and expanded results are much closer, as expected. In principle, the most suitable choice for $\mu_s$ would be $\mu_s\simeq 100\,\MeV$. However, the implementation of the nonperturbative contributions as described in Section~\ref{sec:dirtyWorld} includes all quarks (other than the top). This restricts our choice to $\mu_s \ge m_b$. As for $A_{LR}^\text{vNLO exp}$ shown in Figure~\ref{fig:ALRvirtmu}, the corrections are reduced for smaller scales of $\mu_s$. Furthermore, the dependence on $\mu_s$ is drastically reduced from LO to NLO, with only minor changes at NLO from $\mu_s=30\,\GeV$ to $\mu_s=5\,\GeV$. This gives some confidence that the choice $\mu_s=5\,\GeV$ leads to numerical results that are not unreasonable, even if smaller values of $\mu_s$ are preferred. However, we recall the potentially sizeable impact of NLL corrections. 

\begin{figure}[t]
\centering  
    \includegraphics[width=0.9\textwidth]{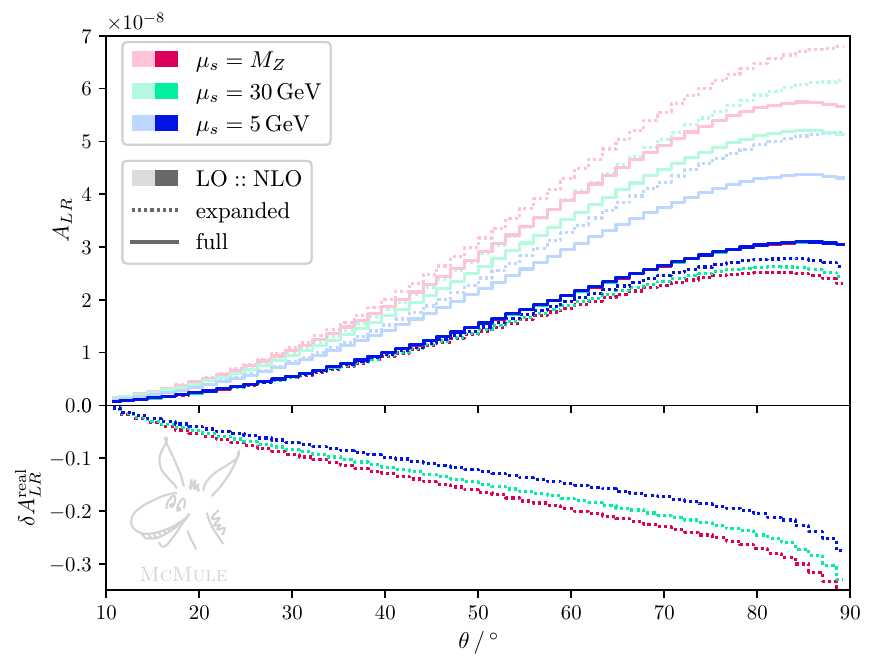}
\\[10pt]
    \caption{
    Same as Figure~\ref{fig:mollercm} but with modified angular cuts $10^\circ \le \theta_{3,4} \le 170^\circ$.
    }
    \label{fig:mollercmLA}
\end{figure}

In the lower panel of Figure~\ref{fig:mollercm} we show the impact of the real corrections through the ratio
\begin{align} \label{eq:deltaReal}
    \delta A_{LR}^\text{real}&\equiv 
    \frac{A_{LR}^\text{NLO exp} - A_{LR}^{\text{vNLO exp}}}{A_{LR}^\text{vNLO exp}}\, .
\end{align}
With the standard cuts on the scattering angles \eqref{cutsThetaCoM} the size of the real corrections ranges between 0 and $-7.5$\%. However, we emphasise that the impact of the real corrections strongly depends on the precise definition of the observable. To illustrate this, in Figure~\ref{fig:mollercmLA} we show the same results, however with modified angular cuts $10^\circ \le \theta_{3,4} \le 170^\circ$ rather than \eqref{cutsThetaCoM}. As can be seen from the lower panel, real corrections now amount to up to $-30$\%. Related to this, also the difference between the full and the expanded version of $A_{LR}$ changes between the two scenarios with different angular cuts. The improved convergence of the perturbative series for small values of $\mu_s$ remains the same. For these results, the NNLO QED corrections are at the permille level and, thus, have no significant impact.

We now turn to some examples in the laboratory frame, where we define $\tilde{E}_h$ and $\tilde{\theta}_h$ as the energy and scattering angle of the more energetic (hard) electron, and similarly $\tilde{E}_s$ and $\tilde{\theta}_s$ for the less energetic (soft) electron. To avoid ambiguities, we reject events with a very small energy difference, i.e. we request $\tilde{E}_h-\tilde{E}_s \ge 10\,\MeV$. 

\begin{figure}[t]
\centering  
    \includegraphics[width=0.9\textwidth]{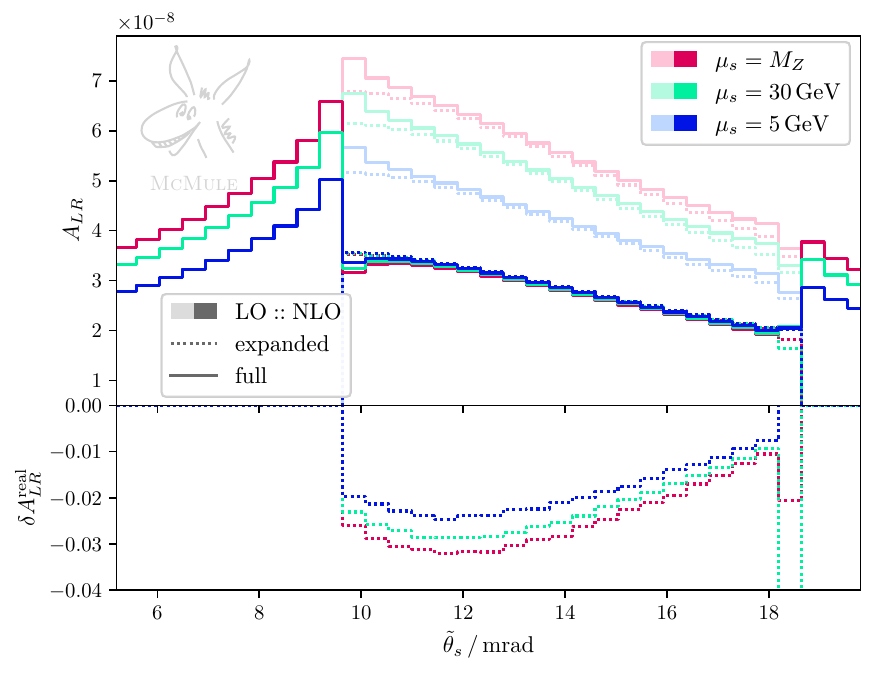}
\\[10pt]
    \caption{$A_{LR}$ for the MOLLER experiment, as a function of $\tilde\theta_s$ in the laboratory frame at LO and NLO in RGE-improved perturbation theory, for three choices of $\mu_s$. Solid and dotted lines correspond to \eqref{ALRfull} and \eqref{ALRexp} respectively.
    The hard cuts around $\tilde{\theta}_c=9.64\,\rm{mrad}$ and $\tilde{\theta}_c=18.58\,\rm{mrad}$ are a consequence of $2\to2$ kinematics and \eqref{cutsThetaCoM}. The bottom panel shows the impact of real correction for the expanded version of $A_{LR}$ for $\tilde\theta_s > \tilde\theta_c$. For $\tilde\theta_s < \tilde\theta_c$, the expanded version \eqref{ALRexp} is not defined. 
    }
    \label{fig:mollerlabsoft}
\end{figure}

\begin{figure}[t]
\centering  
    \includegraphics[width=0.9\textwidth]{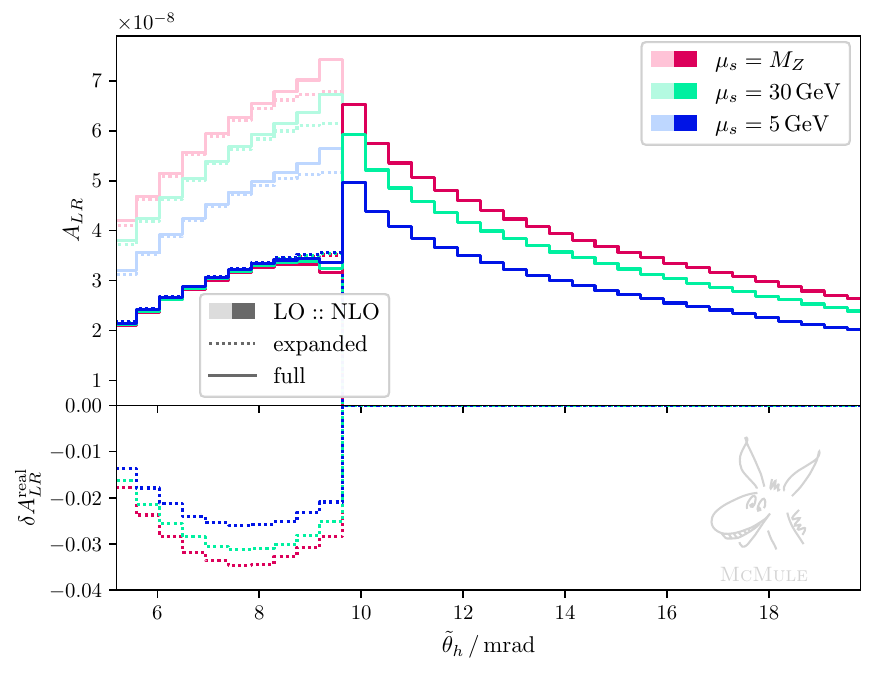}
\\[10pt]
    \caption{$A_{LR}$ for the MOLLER experiment, as a function of $\tilde\theta_h$ in the laboratory frame at LO and NLO in RGE-improved perturbation theory, for three choices of $\mu_s$. Solid and dotted lines correspond to \eqref{ALRfull} and \eqref{ALRexp} respectively.
    The hard cut around $\tilde{\theta}_c=9.64\,\rm{mrad}$ is a consequence of $2\to2$ kinematics.  The bottom panel shows the impact of real correction for the expanded version of $A_{LR}$ for $\tilde\theta_h < \tilde\theta_c$. For $\tilde\theta_h > \tilde\theta_c$, the expanded version \eqref{ALRexp} is not defined. 
    }
    \label{fig:mollerlabhard}
\end{figure}

The angular cuts \eqref{cutsThetaLab} are more or less equivalent (slightly more restrictive) to the corresponding cuts in the centre-of-mass frame \eqref{cutsThetaCoM}.
They translate to $54.84^\circ \leq\theta_{3,4} \leq 128.54^\circ$.
With tree-level kinematics,  there is a lower (upper) limit for the soft (hard) electron at 
\begin{align}\label{def:thc}
\tilde{\theta}_c &\equiv 
\sqrt{\frac{E_\text{beam}^2-m_e^2}{(E_\text{beam}+m_e)^2-4m_e^2}} \simeq 9.64\,\mrad\, .
\end{align}
In addition, due to  the request that both electrons satisfy~\eqref{cutsThetaCoM} and \eqref{cutsThetaLab}, the allowed range of the scattering angles in the laboratory frame is $5\,\mrad\leq\tilde{\theta}_{3,4} \leq 18.58\,\mrad$ for tree-level kinematics.
This is visible in Figures~\ref{fig:mollerlabsoft} and \ref{fig:mollerlabhard}, where we show $A_{LR}$ as a function of $\tilde{\theta}_s$ and $\tilde{\theta}_h$ respectively. In the region that is allowed by ${2}\to{2}$ kinematics, the picture is similar to Figure~\ref{fig:mollercm}, however with larger differences between the full and expanded results for $A_{LR}$. The real corrections are of the order of $-3$\%. However, in the region not accessible by ${2}\to{2}$ kinematics ($\tilde\theta_s < \tilde\theta_c$, $\tilde\theta_s > 18.58\,\mrad$, and $\tilde\theta_h > \tilde\theta_c$) the real corrections are absolutely crucial. In fact, the expanded version $A_{LR}^\text{exp}$ is not even defined in this region, since $\D\sigma^{(0)}=\D\sigma^{(0,1)}=0$. Thus, we can only consider the full version of $A_{LR}$.
In that region, also the impact of NNLO QED corrections is large and reaches up to $10\%$. This is not surprising, since the NNLO QED calculation in fact is a NLO result in this region, due to the kinematic constraints. However, it is a clear illustration of the importance of combining QED with EW corrections to obtain accurate results for the whole range of scattering angles in the laboratory frame. Furthermore, this example illustrates that a fully differential implementation of combined EW and QED corrections to M{\o}ller scattering benefits a direct comparison to experimental results. If required, it is easily possible to add further cuts such as a cut on the energy of the electrons or include an energy spread of the incoming beam. 

\section{Conclusion and outlook} \label{sec:conclusion}

We have presented a fully differential calculation for M{\o}ller scattering at low energies within LEFT, including terms suppressed by $s/\Mew^2$. The large logarithms of the form $\ln(s/\Mew^2)$ are resummed at leading logarithmic accuracy.  This is combined with fixed-order contributions, namely NLO EW effects and NNLO QED effects. The results have been implemented in \mcmule{} which allows to obtain realistic predictions including arbitrary experimental cuts. 

While it is possible to compute any IR safe observable at this order, EW effects are crucial for parity-violating quantities. Hence, for the presentation of our results we have focused on $A_{LR}$. This quantity has received a lot of attention in the literature and it has been known for decades that there are very large NLO corrections. In view of the upcoming MOLLER experiment, an improved evaluation of $A_{LR}$ beyond NLO is therefore highly desirable, and several groups have presented partial results. In our view, using LEFT is the most appropriate approach for such a computation. Indeed, LEFT systematically exploits the hierarchy between the EW and the low scales. This leads to simpler loop integrals. Furthermore, it facilitates the resummation of all logarithms $\ln(s/\Mew^2)$. Given the progress in the determination of the two-loop anomalous dimensions in LEFT, a complete resummation at the next-to-leading logarithmic accuracy appears to be feasible. In fact, we have investigated the potential impact of NLL contributions and they are significant at the percent level. In typical applications of RGE-improved calculations, a NLL resummation is combined with a NLO calculation. However, in view of the large NLO corrections mentioned above, it is desirable to combine NLL RGE with a fixed-order NNLO EW calculation. This requires the two-loop matching of the LEFT to the SM. Furthermore, a two-loop calculation within the EFT is required.

In addition to computational advantages, LEFT also offers the option to systematically investigate new-physics effects induced by heavy degrees of freedom. In fact, the calculation can be done with arbitrary Wilson coefficients, rather than those obtained through matching. The additional complications are minor. This approach also reveals that at NLO, the contribution of non-SM operators vanish in the limit $m_e\to 0$. This will make it challenging to obtain strong limits on the Wilson coefficients of such operators.

A complication that arises in the computation of $A_{LR}$ for very low energies are the non\-perturbative hadronic contributions. In pure QED, up to NNLO this can be dealt with by using experimental or lattice input for $\Pi_{\gamma\gamma}$.  For the EW contribution, also
$\Pi_{\gamma{Z}}$ is required, for which there are phenomenological approaches. 
However, a proper EFT treatment would require to completely disentangle the perturbative and nonperturbative regimes. This requires to match another EFT with hadronic degrees of freedom nonperturbatively to LEFT. A natural scale for such a matching would be around $2\,\GeV$ which, unfortunately, is rather high for chiral perturbation theory. While this is a serious complication, it is not unique to parity-violating electron scattering. The combination of LEFT with nonperturbative effects is a topic that is actively under investigation~\cite{Tomalak:2020zfh,Aebischer:2021uvt,Cirigliano:2023fnz,Cirigliano:2024nfi}.

In addition to improving the theoretical description of $A_{LR}$ for the MOLLER experiment, the inclusion of LEFT into the \mcmule{} framework also allows for extensions to other energies and processes. While the use of LEFT prevents a direct evaluation of $A_{LR}$ at EW energies or above, it is still perfectly possible to use it for electron-positron colliders up to $\sqrt{s}\simeq 10\,\GeV$. This includes Bhabha scattering and muon- or tau-pair production and muon-electron scattering. EW corrections to such processes at low energies have been considered before~\cite{Alacevich:2018vez, Arbuzov:2022mij, Kollatzsch:2022bqa}. However, typically this was done through a SM calculation. Generally speaking, the inclusion of EW effects to low-energy leptonic processes is efficiently done with the setup presented in this work. There is also a flexibility to study different input schemes or renormalisation schemes, if required. A more delicate extension is to consider processes with external hadrons. Indeed, a sophisticated treatment of hadronic effects becomes even more important for extensions to parity-violating electron scattering off hadrons, as will be done by P2. 

Effective theories have been widely advocated as systematic and efficient tools to deal with complex physical situations with a hierarchy of scales. We view the work presented in this article as a starting point to use this tool and extend \mcmule{} to include EW effects to low-energy processes in a fully differential way.

\subsection*{Acknowledgement}

We would like to thank Yannick Ulrich for his indispensable support with technical aspects related to \mcmule. We gratefully acknowledge the help of Jason Aebischer, Pol Morell, Marko Pesut, and Javier Virto, for their help with the two-loop anomalous dimensions and the use of {\tt DsixTools}. It is a pleasure to thank Martin Hoferichter, Frederic No\"el, Peter Stoffer, and  Dominik St\"ockinger for very useful discussions regarding the HVP, the matching process, and input schemes.
SK and DR acknowledge support by the Swiss National Science Foundation (SNSF) under grant 207386.
DM has received funding from the European Union’s Horizon 2020 research and innovation programme under the Marie Sk{\l}odowska-Curie grant agreement No 884104 (PSI-FELLOW-III-3i).

\appendix

\section{Input parameters} \label{sec:appendixinput}

The EW scale defines our starting point for the RGE and is fixed to the on-shell $Z$ boson mass
\begin{displaymath}
    \Mew =91.1880\,\text{GeV}\,.
\end{displaymath}
The LEFT parameters at the EW scale, obtained by matching the LEFT onto the SM up to one-loop accuracy, are taken from \cite{Dekens:2019ept}. All SM parameters appearing in the matching conditions are renormalised in the pure $\overline{\text{MS}}$ scheme. We derive their values at $\mu=\Mew$ using the input parameters defined in {\tt DsixTools} \cite{Fuentes-Martin:2020zaz,Celis:2017hod}. The couplings and heavy masses are given by
\begin{align}
\label{eq:inputval}
\begin{split}
\begin{aligned}
    \bar{\alpha}&=1/127.90\,, & \alpha_s&=0.11850\,, & \hat{s}_W^2&=0.23151\,, \\
    M_Z&=91.486\,\text{GeV}\,, & M_H&=130.60\,\text{GeV}\,, & M_t&=169.00\,\text{GeV}\,.
\end{aligned}
\end{split}
\end{align}
After diagonalising the up-quark Yukawa matrix to arrive at the up-basis used in \cite{Dekens:2019ept}, and rephasing the unitary transformations to reach the standard phase convention, we obtain the CKM matrix
\begin{displaymath}
    V_{\text{CKM}}=\begin{pmatrix}
    0.9744-0.0174\,i & 0.2243-0.0039\,i & 0.0010-0.0033\,i \\
    -0.2241+0.0040\,i & 0.9735-0.0174\,i & 0.0422 \\
    0.0084-0.0033\,i & -0.0414 & 0.9991
    \end{pmatrix}\,.
\end{displaymath}
The light scales explicitly appear only in the matching conditions and RGEs for LEFT parameters of dimension less than 6. While their running is taken into account in the solutions of the RGEs for higher-dimensional operators, we choose to employ the usual PDG values~\cite{ParticleDataGroup:2024cfk} for the masses of the light fermions when evaluating the matrix elements in regimes where RGE-improved perturbation theory is applicable.
We use
\begin{align}
\begin{split}
\begin{aligned}
    m_e&=0.510998950\,\text{MeV}\,, & m_{\mu}&=105.658375\,, & m_{\tau}&=1.77686\,\text{GeV}\,, \\
    m_d&=4.7\,\text{MeV}\,, & m_s&=93.5\,\text{MeV}\,, & m_b&=4.183\,\text{GeV}\,, \\
    m_u&=2.16\,\text{MeV}\,, & m_c&=1.2730\,\text{GeV}\,.
\end{aligned}
\end{split}
\end{align}
The precise value of the light quark masses has no numerical impact on the results.
The left-handed neutrinos in the SM are assumed to be massless, i.e. the PMNS matrix is set to the unit matrix.

For the results presented in Section~\ref{sec:ResMoller}, we have used the on-shell scheme for $\alpha$ with
\begin{align}
    \alpha = 1/137.03599908\,.
\end{align}
\section{Explicit results}\label{sec:appendixexplicit}
In this appendix, we present some explicit results for $A_{LR}$ in M{\o}ller scattering within LEFT, assuming the applicability of perturbation theory and expanding the defining equation of $A_{LR}$ up to dimension 6 and one-loop order.

The tree-level contributions generated by the interference $[\text{QED 0}]\otimes[\text{EW 0}]$, combined with the loop-induced dipole corrections $[\text{QED 0}]\otimes[\text{EW 1}]$, with full dependence on $m_e$, are given by
\begin{align}
     A_{LR}^{\text{LO exp}} &= \frac{F_V\,\bigl(C_{ee,1111}^{V,LL}-C_{ee,1111}^{V,RR}\bigr)+F_S\,\bigl(C_{ee,1111}^{S,RR\,*}-C_{ee,1111}^{S,RR}\bigr)+F_D\,\bigl(C_{e\gamma,11}^*-C_{e\gamma,11}\bigr)\,e\,Q_e}{\sqrt{1-\frac{4m_e^2}{s}}\,\bigl[(t^2 + t u + u^2)^2 - 4\,(t^3 + u^3)\,m_{e}^2 + 
 4\,(t^2 - t u + u^2)\,m_{e}^4\bigr]\,e^2\,Q_e^2}\,,
\end{align}
with
\begin{subequations}
\begin{align}
    F_V&=2\,t\,u\,((t + u)^3 - 2 (t^2 + u^2) m_{e}^2)\,, \\
    F_S&=-2\,t^2\,u^2\,m_{e}^2\,, \\
    F_D&=2\,t\,u\,m_{e}\,((t - u)^2 - 2 (t + u) m_{e}^2)\,.
\end{align}
\end{subequations}
The corresponding analytic expression for $A_{LR}^{\text{NLO exp}}$ with full dependence on $m_e$ is given in repository linked in Section~\ref{sec:perturbative}.

The results given in~\eqref{eq:ALRexplicit} are expanded for $m_e^2\ll s$ and the functions $F_{ij}(y)$ introduced there are defined in terms of the kinematic variable $y=-t/s$. The indices indicate the diagram classes (see Figure \ref{fig:diag}) to which the respective contributions belong, i.~e. $F_{ij}$ stands for the contribution coming from the interference of diagrams $[{\rm QED}\, i] \otimes [{\rm EW}\, j]$. At LO we have
\begin{align}
	F_{00}(y)=\frac{y\,(y-1)}{(1-y+y^2)^2}\,.
\end{align}
The purely bosonic corrections $[\text{QED 1}]\otimes[\text{EW 0}]$ and $[\text{QED 0}]\otimes[\text{EW 2}]$, corresponding to $\gamma\gamma$ and $\gamma Z$ box diagrams in the SM, respectively, are given by
\begin{align}
    F^{\Box}(y)&=F_{10}(y)+F_{02}(y) \nonumber \\
	&=\frac{F_{00}(y)}{(1-y+y^2)^2}\,
		\biggl[2\,(2b-11)\,(1 - y + y^2)^2 - \pi^2\,\bigl(2 - 4 y + 11 y^3 - 13 y^4 + 9 y^5 - 3 y^6\bigr) \nonumber\\
			&\hspace{84pt}+ \Bigl(2\,(1 - y)\,(3 - 3 y + 4 y^3 - 3 y^4) \nonumber \\
			&\hspace{84pt}\hspace{17.7pt}- (1 - y)\,\bigl(2 - 2 y - 7 y^2 + 10 y^3 - 8 y^4 + 3 y^5\bigr)\ln(1-y) \nonumber\\
			&\hspace{84pt}\hspace{17.7pt}+ 2\,\bigl(2 - 4 y + 11 y^3 - 13 y^4 + 9 y^5 - 3 y^6\bigr)\ln y\Bigr)\ln(1-y) \nonumber \\
			&\hspace{84pt}+ \Bigl(2 y\,\bigl(1 + 3 y - 6 y^2 + 8 y^3 - 3 y^4\bigr) \nonumber\\
			&\hspace{84pt}\hspace{17.7pt}+ y\,\bigl(2 - 3 y - 5 y^2 + 8 y^3 - 7 y^4 + 3 y^5\bigr)\ln y\Bigr)\ln y  \nonumber \\
			&\hspace{84pt}- 6\,(1 - y + y^2)^2\,\ln\left(\frac{\mu_s^2}{s}\right)
			\biggr]\,.
\end{align}
The coefficient $b$ is related to our definition of evanescent operators for Chisholm identities, which agrees with the conventions adopted in \cite{Dekens:2019ept}. It also enters the matching conditions for the Wilson coefficients at one loop. While different values define different schemes, the final results remain independent of the choice of $b$. Contributions due to penguin diagrams, $[\text{QED 0}]\otimes[\text{EW 3}]$, lead to
\begin{align}
    F_{03}^{\triangle}(y)&=-\frac{2\,F_{00}(y)}{9}\,\biggl[4-3\ln\bigl((1-y)\,y\bigr)
    +6\ln\left(\frac{\mu_s^2}{s}\right)\biggr]\,, \\
    F_{03}(y,m)&=-\frac{F_{00}(y)}{3}\,\biggl[\frac{10 s\,(1 - y)\,y - 12 m^2}{3 s\,(1 - y)\,y}+2\ln\left(\frac{\mu_s^2}{m^2}\right) + \biggl(1-\frac{2m^2}{s\,y}\biggr)\Real B(-s\,y,m,m) \nonumber\\
		&\hspace{61.2pt}+\biggl(1-\frac{2m^2}{s\,(1-y)}\biggr)\Real B(-s\,(1-y),m,m) \biggr]\,,\label{eq:F03}
\end{align}
where $F_{03}^{\triangle}$ describes corrections to the electron-photon vertex involving a $Z$ boson in the SM, while $F_{03}$ captures closed fermion loops. The electron loop contributions can be obtained from $F_{03}$ in the massless limit (the factor 2 is due to the fact that $C_{ee,1111}^{V,LL}=2C_{ee,11pp}^{V,LL}$ for $p\neq1$). The additional term $\frac{4}{3}(C_{ee,11pp}^{V,LL}-C_{ee,11pp}^{V,RR})F_{00}(y)$ in the fourth line of \eqref{eq:ALRvirt} explicitly breaks the symmetry of penguin-diagram-induced lepton-loop contributions under the change of flavour $p\leftrightarrow r$, for \mbox{$p,r\in\{1,2,3\}$}. It emerges upon applying the Fierz identities in the $(\overline{L}L)(\overline{L}L)$ and $(\overline{R}R)(\overline{R}R)$ sectors, which already give rise to evanescent operators at tree level in the matching of the LEFT onto the SM. Once the evanescent operators are properly taken into account at the given loop order, as described in \cite{Dekens:2019ept}, this term cancels against the additional finite renormalisation of physical operators entering the matching results for $C_{ee,1111}^{V,LL}$ and $C_{ee,1111}^{V,RR}$. In the end, when expressing the LEFT coefficients and parameters in terms of SM parameters, the flavour symmetry of closed fermion loop contributions is restored. The function $B$ in \eqref{eq:F03} denotes the {\tt DiscB} function in {\tt Package-X},
\begin{equation}
B(x,m,m) = \frac{\sqrt{x\,(x-4 m^2)}}{x}\,\ln \left(\frac{\sqrt{x\,(x-4 m^2)}+2 m^2-x}{2 m^2}\right)\,,
\end{equation}
containing the normal threshold discontinuity of the Passarino-Veltman $B_0$ function. Finally,
\begin{align}
	F_{20}(y,m)&=\frac{4\,F_{00}(y)}{3\,(1 - y + y^2)}\,\biggl[\frac{5s\,y\,(1 - 2 y + 2 y^2 - y^3)-12 (1 - 2 y + 4 y^3 - 2 y^4)\,m^2}{3s\,(1 - y)\,y} \nonumber\\
		&\hspace{86.6pt}+ (2 - y)\,y^2\,\biggl(1-\frac{2m^2}{s\,(1-y)}\biggr)\Real B(-s\,(1-y),m,m) \nonumber \\
		&\hspace{86.6pt}+ (1 + y)\,(1 - y)^2\,\biggl(1-\frac{2m^2}{s\,y}\biggr)\Real B(-s\,y,m,m) \nonumber \\
		&\hspace{86.6pt}+ \zeta\,(1 - y + y^2)\,\ln\left(\frac{\mu_s^2}{m^2}\right)\biggr]
        \label{eq:F20}
\end{align}
describes all corrections $[\text{QED 2}]\otimes[\text{EW 0}]$ due to VP diagrams in QED. As before, electron-loop contributions are obtained by taking the massless limit. The parameter $\zeta$ accounts for the change in the renormalisation scheme of $\alpha$. The $\overline{\text{MS}}$ scheme is obtained by setting $\zeta=1$, while $\zeta=0$ corresponds to the on-shell scheme.

\bibliographystyle{JHEP}
\bibliography{ew-matching}

\end{document}